\documentclass[
 conference
]{IEEEtran}
\IEEEoverridecommandlockouts

\ifCLASSINFOpdf
\else
\fi
\usepackage[cmex10]{amsmath}

\usepackage{cite}
\usepackage{enumerate}
\usepackage{bbm}
\usepackage{amsmath}
\usepackage{amsfonts}
\usepackage{graphicx}
\usepackage{epstopdf}
\usepackage{caption2}
\usepackage{subfigure}
\begin{document}
	%
	\title{Price Competition with LTE-U and WiFi }
	
	\author{\IEEEauthorblockN{Xu Wang and Randall A. Berry}
	\IEEEauthorblockA{ECE Department, Northwestern University, Evanston, IL 60208\\
			Email: xuwang2019@u.northwestern.edu, rberry@ece.northwestern.edu}
		\thanks{This research was supported in part by NSF grants TWC-1314620, AST-1547328 and CNS-1701921.  Some results in the paper were presented at IEEE INFOCOM, Paris, France, April 2019 \cite{wang2019price}.}
	}


	\maketitle

	\begin{abstract}
	LTE-U is an extension of the Long Term Evolution (LTE) standard for operation in unlicensed spectrum. LTE-U differs from WiFi, the predominant technology used in unlicensed spectrum in that it utilizes a duty cycle mode for accessing the spectrum and allows for a more seamless integration with LTE deployments in licensed spectrum. There have been a number of technical studies on the co-existence of LTE-U and WiFi in unlicensed spectrum In this paper, we instead investigate the impact of such a technology from an economic perspective. We consider a model in which an incumbent service provider (SP) deploys a duty cycle-based technology like LTE-U in an unlicensed band along with operating in a licensed band and competes with one or more entrants that only operate in the unlicensed band using a different technology like WiFi.  We characterize the impact of a technology like LTE-U on the market outcome and show that the welfare impacts of this technology are subtle, depending in part on the amount of unlicensed spectrum and number of entrants. The difference in spectral efficiency between LTE and WiFi also plays a role in the competition among SPs. Finally, we investigate the impact of the duty cycle and the portion of unlicensed spectrum used by the technology. 
	\end{abstract}

	\IEEEpeerreviewmaketitle

	\newtheorem{definition}{Definition}
	\newtheorem{proposition}{Proposition}[section]
	\newtheorem{corollary}{Corollary}[section]
	\newtheorem{theorem}{Theorem}[section]
	\newtheorem{lemma}{Lemma}[section]
	\section{Introduction}
	
Offloading traffic to unlicensed spectrum has been a vital approach for wireless service providers (SPs)  to meet the ever rising demand for mobile data and retain control over profit margins \cite{jia2008bandwidth,kavurmacioglu2012competition,ghosh2013quality}. Indeed, in 2016, there was more mobile data traffic offloaded to unlicensed bands than served in licensed spectrum worldwide \cite{cisco17}.  These trends are expected to continue with 5G \cite{mekuria2019spectrum,ali2018resource} and have led to the development of technologies for unlicensed access that are based on the LTE technology that SPs utilize in licensed spectrum. The two main examples of this are LTE in unlicensed spectrum (LTE-U) and License Assisted Access (LAA).  These differ in several ways from the WiFi technologies that are widely used in the same unlicensed spectrum. For example, both LTE-U and LAA utilize LTE's carrier aggregation capability to essentially combine a SPs licensed and unlicensed spectrum. Moreover, LTE-U differs in that it does not employ a listen-before-talk (LBT) protocol as used by WiFi, but instead is based on a duty-cycle based approach.\footnote{LAA does employ LBT, which is required in some parts of the world. LTE-U was developed first and is being deployed in countries where LBT is not required for unlicensed channel access. For example, T-Mobile launched LTE-U in spring 2017 \cite{Tmobile} to serve customers in many cities in the US. }  This led to much interest in studying the co-existence of WiFi and LTE-U from a technical point of view, e.g.~\cite{maglogiannis2019enhancing, gawlowicz2018enabling, leanh2019orchestrating,manzoor2018ruin,bairagi2018bargaining,elgendi2018break, cano2015coexistence,sriyananda2016multi,almeida2013enabling,hamidouche2016multi}.
In this paper, we instead examine the impact of such technology from market point-of-view. Namely, we seek to understand the impact of a SP using a technology like LTE-U on the competition with other SPs that utilize a technology like WiFi.

 We consider a scenario similar to that in \cite{nguyen2011impact,nguyen2015free}, where  SPs compete for customers by announcing prices for their service (see also \cite{maille2009price,liu2014competitionopen,bairagi2018game,zhu2018contracts,ghosh2019tiered,hamidouche2019contract}). The customers select SPs based on the sum of the price they pay for service and a congestion cost that is incurred for using the given band of spectrum.  In \cite{nguyen2011impact,nguyen2015free}, the SPs compete by announcing one price for service in an unlicensed band and a different price for service in any licensed band that the SP may own. When the LTE-U technology is adopted, we instead assume the SP can announce a single price because of the seamless integration of LTE technology on both the licensed and unlicensed bands. In our model, the duty cycle mechanism of LTE-U is considered. Under the duty cycle setting, the incumbent SP (with licensed spectrum) is able to use both its licensed spectrum and a portion of the unlicensed spectrum to serve customers when the duty cycle is `ON' while it can only use its own licensed band when the duty cycle is `OFF'. In such a scenario, we model customers as being sensitive to the average congestion they experience across the whole duty cycle.  In addition, LTE on unlicensed spectrum can provide higher spectrum efficiency than WiFi systems \cite{al20155g}. This is because LTE is a synchronous system and adopts scheduling-based channel access instead of contention-based random access \cite{zhang2015lte}. In our model, we use one spectrum efficiency factor to capture the difference while ignoring other complicated technical differences between LTE and WiFi.

    We use $\alpha$ and $\beta$ to denote the duty cycle and the portion of unlicensed spectrum that are used for LTE-U, respectively.  We first consider $\alpha$ and $\beta$ as fixed parameters, e.g. determined by a regulator. For example, currently LTE-U channel bandwidth is set to 20 MHz which corresponds to the smallest channel width in WiFi and Qualcomm recommends that LTE-U should use a period of 40, 80 or 160 ms, and limits the  maximal duty cycle to $50\%$ \cite{gawlowicz2017ltfi}. We show that in the monopoly market, using LTE-U hurts the SP's revenue and social welfare. We then show that when there are multiple entrant SPs in the market, adopting LTE-U technology can help the incumbent SP to increase revenue and also benefit social welfare when the bandwidth of unlicensed spectrum is small. When there is only one entrant SP in the market, we show that it is possible for LTE-U technology to hurt the revenue of the incumbent. We also investigate the impact of different spectral efficiency between LTE and WiFi. We show that if the efficiency advantage of LTE over WiFi is large, LTE-U may benefit the incumbent's revenue and customer welfare in different competition scenarios.  
    
     Then we consider $\alpha$ as a controllable parameter with fixed $\beta$. We show that with multiple entrant competitors, the incumbent's revenue  increases with $\alpha$ while with one entrant, the incumbent may prefer a small $\alpha$. Finally, we consider varying $\alpha$ and $\beta$ while keeping the utilization of unlicensed spectrum ($\alpha\beta$) constant. We show that when the unlicensed bandwidth is small, the incumbent may prefer lower $\alpha$ and higher $\beta$. But when the unlicensed bandwidth is large, the incumbent may prefer a higher $\alpha$ and a lower $\beta$.
 
	In terms of other related work,  \cite{bykowsky2016economic} also considers an economic model of LTE-U and WiFi. In  \cite{bykowsky2016economic} the focus is not on competition between LTE-U and WiFi providers (there is only one licensed service provider) but rather on understanding how LTE-U impacts the service selection of a finite number of users, each with a ``congestion tolerance" for the service they select. In this work, WiFi is a free option that is congestible, while the licensed service is not congestible but is available at a cost. In  \cite{yu2016auction}, authors propose an auction based spectrum sharing framework to investigate the possibility of cooperation between LTE and Wi-Fi in the unlicensed band. The proposed mechanism make SPs to explore the potential benefits of cooperation before deciding whether to enter head-to-head competition.	In \cite{wang2018impact}, the authors mainly focus on competition instead of cooperation. They analyze the market impact when the incumbent SP offers a bundle price for service on licensed and unlicensed band to compete with entrant SPs.  Different from bundling services, this paper focus on the bundling of spectrum, because LTE-U aggregates the spectrum directly, which is able to provide more seamless service to customers.

	The rest of the paper is organized as follows. Our model is described in Section \ref{sec:model}.  We first consider the monopoly case in Section \ref{sec:mono}. Then we treat $\alpha$ and $\beta$ as fixed parameters in Section \ref{sec:cases} and compare the results with the monopoly case and the model in \cite{nguyen2011impact,nguyen2015free}. In Section \ref{sec:spectral_efficiency}, we look at how different spectral efficiencies impact the competition in the market. In Section \ref{sec:alpha}, we view $\alpha$ and $\beta$ as controllable variables  and  investigate their impact on the incumbent SP. Some numerical results are shown in Section \ref{sec:numerical}. Finally, we conclude in Section \ref{sec:conclusion}.

	\section{System model}
	\label{sec:model}
	We consider a market with one incumbent SP  and $N$ entrant SPs, where the incumbent SP uses a duty-cycle based technology such as LTE-U. In the following we will simply refer to this as LTE-U, though as noted previously this is not intended to model every aspect of LTE-U. The incumbent SP is assumed to  possess its own licensed band of spectrum with bandwidth $B$, while entrants have no licensed spectrum. There is a single unlicensed band with bandwidth $W$ that can be used by both the incumbent and entrant SPs. When the incumbent SP applies the LTE-U technology, it uses carrier aggregation on the unlicensed band and operates in a duty cycle mode.  When LTE-U is in `ON' mode, we assume that it is always using a portion of the unlicensed spectrum so that entrant SPs are not able to operate over this portion (e.g.  due to LBT, the entrants would sense the incumbents presence and not transmit). We use $\alpha$ to denote the percentage of time that the SP aggregates the unlicensed spectrum. We use $\beta$ to denote percentage of unlicensed spectrum that the incumbent uses when the duty cycle is in `ON' state, i.e., when LTE-U is 'ON', the bandwidth that the incumbent can use becomes $B+\beta W$.

	The SPs are assumed to compete for a common pool of infinitesimal customers by setting prices for their services.  Without loss of generality, we assume that the incumbent is SP 1, and all the entrant SPs  are indexed from 2 to $N+1$.  The price announced by SP $i$ is denoted by $p_i$. The SPs  serve all customers that accept their price.  The revenue of SP $i$ is then $x_ip_i$, where $x_i$ is the customer mass that accept price $p_i$.  
	
	As in \cite{maille2009price,liu2014competitionopen,wang2018impact}, a SP's service is characterized by a congestion cost. The congestion that the customers experience in a band is denoted by $g(X,Y)$, which is assumed to be increasing in the total customer mass $X$ on the band and decreasing in the  bandwidth used  $Y$. Here, we assume a specific form $g(\frac{X}{Y})$, where $g(\cdot)$ is  a convex increasing function with $g(0)=0$ and $\frac{X}{Y}$ is the number of users per unit bandwidth.  When the incumbent SP applies LTE-U technology,  the congestion that the customers experience will vary across the duty cycle. We assume that customers are sensitive to the average congestion across the duty cycle.\footnote{This is reasonable as over the time-scale that customer select SPs they will receive service over many duty cycles.}  The average congestion of customers served by the incumbent SP is then given by  
	\begin{equation}	
	\hat{g}_{in}{(x_1)}= \alpha g\left(\frac{x_1}{B+\beta W}\right)+(1-\alpha) g\left(\frac{x_1}{B}\right).\nonumber
	\end{equation}  
	The average congestion experienced by customers who choose an entrant SPs is
	 \begin{equation}
	\hat{g}_{en}\mathbf{(x)} = \alpha g\left(\frac{\sum\limits_{j=2}^{N+1} x_j}{(1-\beta) W}\right)+(1-\alpha) g\left(\frac{\sum\limits_{j=2}^{N+1} x_j}{W}\right).\nonumber
	\end{equation} 	
	Note that as in \cite{nguyen2011impact}\cite{nguyen2015free}, the congestion experienced in the unlicensed band by a customer of an entrant SP is the same for all entrants and depends on the total traffic across all entrants.  This is modeling the fact that the entrants are all employing a technology like WiFi to share this band.   Also note that we assume that when the  LTE-U  duty cycle is on, the entrant can only use the remaining $(1-\beta)W$ of the spectrum. 
	
	As in \cite{nguyen2011impact,nguyen2015free}, we assume that customers seek to receive service from the SP with the lowest {\it delivered price}, which is given by the sum of the announced price and the average congestion cost of that SP's service. This captures the fact that customers are sensitive both to cost of service and the quality of service.  Hence, for the incumbent SP, the delivered price $d_1(p_1,{\bf x})$ is denoted by
	$p_1 + \hat{g}_{in}{(x_1)}$. For an entrant SP $i$, $i \ge 2$, its delivered price $d_i(p_i,{\bf x})$ is given by $p_i +\hat{g}_{en}\mathbf{(x)}$.
	
	We assume that customers are characterized by an inverse demand function $P(q)$, which indicates the delivered price at which a mass of $q$ customers are willing to pay for service.  As in much of the prior literature, we assume  $P(q)$ is concave decreasing. Each customer is infinitesimal so that a single customer has a negligible effect on the congestion in any band. Therefore, given the announced price by the SPs, the demand of service for each SP $i$ is assumed to satisfy the Wardrop equilibrium conditions \cite{wardrop1952some}. In our model, the conditions for the SPs are 
		\begin{eqnarray}
		\label{eqn:wardrop1}
		&&d_i(p_i,{\bf x}) = P\left(\sum\limits_{j=1}^{N+1} x_j\right),{\rm for}  \;  x_i>0,\nonumber\\
		&&d_i(p_i,{\bf x}) \ge P\left(\sum\limits_{j=1}^{N+1} x_j\right),{\rm for} \;\forall i. \nonumber
		\end{eqnarray}
	The conditions imply that at the Wardrop equilibrium, all the SPs serving a positive amount of customers will end up with the same delivered price, which is given by the inverse demand function. A Nash equilibrium of the game is one in which the customers are in a Wardrop equilibrium and no SP can improve their revenue by changing their announced price (anticipating the impact this will have of the Wardrop equilibrium).
	
	At an equilibrium, the customer surplus is defined as the difference between the delivered price each customer pays and the amount it is willing to pay, integrated over all the customers, i.e.,  
		\begin{equation}
		\label{eqn:csdefinition}
		CS = \int_{0}^{Q}P(q)-P(Q)dq,
		\end{equation}
		where $Q = \sum\limits_j x_j$. The social welfare of the market is the sum of consumer welfare and the SPs' revenue:
		\begin{equation}
		\label{eqn:swdefinition}
		SW = CS+ \sum\limits_j p_jx_j.
		\end{equation}

	\section{Monopoly scenario}
	\label{sec:mono}
	We  first examine a scenario in which there is only a single incumbent and no entrants.  Hence, the incumbent is a monopolist and can use both the licensed and unlicensed band.  Our goal in this section is to show that for our LTE-U model, such a monopolist would have no incentive to deploy the new technology.  This shows that in later sections when the incumbent does deploy such a technology that it is due to competitive factors and not an inherent advantage of the technology.

	In this section, we allow the incumbent to offer both service using LTE-U (with a given $\alpha$ and $\beta$)  and additionally an "unlicensed service" that uses the remainder of the unlicensed band when the LTE-U duty cycle is off.\footnote{In subsequent sections, the incumbent will only offer service using LTE-U or unlicensed service, instead of this combination.} This ensures that using LTE-U does not reduce the amount of spectrum the incumbent has access to.  The incumbent's revenue optimization is then given by: 
			\begin{eqnarray}
			\label{eqn:mono_optimization}
			&\max_{p_1^l,p_1^u} \;\;\; p_1^lx_1^l +p_1^ux_1^u\\
			{\rm s.t.} & p_1^l +  \alpha g\left(\frac{x_1^l}{B+\beta W}\right)+(1-\alpha) g\left(\frac{x_1^l}{B}\right) = P(x_1^l+x_1^u),\nonumber\\
			& p_1^u +  \alpha g\left(\frac{x_1^u}{(1-\beta) W}\right)+(1-\alpha) g\left(\frac{x_1^u}{W}\right) = P(x_1^l+x_1^u),\nonumber\\
			&  p_1^l,p_1^u\ge0. \nonumber
			\end{eqnarray}	
	Here, $p^l_1$ is the price the incumbent offers for serving $x_1^l$ customers using LTE-U; $p^u_1$ and $x^u_1$ are the corresponding values for the unlicensed service.  The first two constraints enforce the Wardrop equilibrium conditions for these two services. Note also that if we set $\alpha = 0$ and $\beta = 0$, then this reduces to a model as in \cite{nguyen2011impact} in which the incumbent does not employ LTE-U and offers separate licensed and unlicensed service.   
		
	Consider the expected congestion for the LTE-U service in (\ref{eqn:mono_optimization}). Given the convexity of the congestion function $g(\cdot)$, we have the following inequality:
		\begin{eqnarray}
		\label{eqn:congestion}
		\hat{g}_{in}(x_1^l) &\ge& g\left(\frac{\alpha x_1^l}{B+\beta W}+ (1-\alpha)\frac{x_1^l}{B}\right)\nonumber\\
		& \buildrel \Delta \over =& g\left(\frac{x_1^l}{B_e}\right),\nonumber
		\end{eqnarray}
	where $B_e$ denotes the {\it equivalent licensed bandwidth} given by
	\begin{equation}
	\label{eqn:Be}
	B_e = B+\frac{\alpha\beta W}{1+\beta(1-\alpha)W/B}.
	\end{equation}
	Similarly, considering the congestion for the unlicensed service, we have the {\it equivalent unlicensed bandwidth} $W_e$ given by
	\begin{equation}
	\label{eqn:We}
	W_e = W -\frac{\alpha\beta W}{1-\beta(1-\alpha)}.
	\end{equation}
	Note that the congestion is no smaller than in a setting where the incumbent offered separate licensed and unlicensed services (without LTE-U) using the equivalent bandwidth and equality holds when the congestion function $g(\cdot)$ is linear.  Based on this, we have the following  result. 		
	\begin{theorem}
		\label{thm:mono}
		In a monopoly scenario, the incumbent SP can gain no additional revenue by using the LTE-U technology.
	\end{theorem}

	The detailed proof is  in Appendix \ref{append:thm:mono}. Notice that the equivalent licensed bandwidth  increases and the equivalent  unlicensed band decreases but the total amount of equivalent bandwidth in (\ref{eqn:Be}) and (\ref{eqn:We}) decreases when adopting LTE-U. This is the reason that prevents the incumbent from adopting LTE-U. 
	
	\begin{theorem}
		\label{thm:mono_welfare}
		In a monopoly scenario, both customer surplus can social welfare decreases when LTE-U is adopted. 
	\end{theorem}
	The detailed proof is in Appendix \ref{appendix:proof_mono_welfare}. As a result of the loss of equivalent bandwidth, customer surplus and social welfare also decrease when LTE-U is used. Note that in this section, we assume the spectral efficiency of LTE and WiFi are the same. In Section \ref{sec:spectral_efficiency}, we will see that different results emerge when differences in spectral efficiency are accounted for.

	\section{Competition with Fixed $\alpha$ and $\beta$ }
	\label{sec:cases}
	We now  study the case where there is competition between the incumbent and one or more entrants.  We first consider the case where there are multiple entrants and then consider the special case of one entrant.  In both cases we will see that unlike the previous section, the incumbent may now have an incentive to deploy LTE-U. Throughout this section we assume that $\alpha$ and $\beta$ are fixed.  
	
	\subsection{One incumbent \& multiple entrants}
	\label{sec:multiple_entrant}
	Initially, we  assume that there $N\geq 2$ entrants.  Hence, these entrants will compete with each other as well as the incumbent.  The presence of this competition yields the following result on the entrants' equilibrium prices. 

	\begin{lemma}
		\label{lemma:equ}
		 If there are at least two entrant SPs in the market, in any Nash equilibrium every entrant SP $i$ serving a positive mass of customers must have $p_i = 0$ and at least two SPs must announce this price.  
	\end{lemma}
	Lemma \ref{lemma:equ} is similar to a result in \cite{nguyen2011impact} where firms compete in unlicensed spectrum without LTE-U, and so we omit the proof. Essentially, since the entrant SPs are all offering the same service due to sharing  the same spectrum, they will be incentivized to compete the price for this service to zero.  Hence, all the entrant SPs  get no revenue regardless of the incumbent's actions.\footnote{Note here we are ignoring any cost for offering service. If such a cost was included the result would be that the price is competed down to cost, still yielding zero profit.} A corollary of this result is that the incumbent would have no incentive to offer a separate unlicensed service in this setting as its price for this would also be zero.   Based on Lemma \ref{lemma:equ} we give the following result on the incumbent's revenue.
	\begin{theorem}
		\label{thm:one_multiple_profit}
		Consider one incumbent and multiple entrants. Given a fixed $\alpha >0$ and $\beta >0$,  the following hold:
		\begin{enumerate}
			\item The incumbent SP announces a higher price and attracts more customers when LTE-U is adopted. As a result the incumbent SP gets a higher revenue.
			\item The customer mass served by the entrant SPs decreases when the incumbent SP uses LTE-U technology. 
			\item The total customer mass served by the incumbent and entrant SPs is lower when LTE-U is adopted. 
		\end{enumerate}		
	\end{theorem}

 The detailed proof is in Appendix \ref{appendix:proof_one_multiple_profit}. As in the previous section, the use of LTE-U increases the equivalent licensed bandwidth. However,  now that there are multiple entrants keeping the price on the unlicensed band zero, this benefit to the incumbent is not offset with a loss due to the reduced equivalent unlicensed bandwidth.   Due to the improved service by adopting LTE-U, the incumbent is able to announce a higher price and at the same time attract more customers. This leads to an increase in the revenue. However the delivered price will increase, resulting in fewer customers served. 
	
	\begin{theorem}
		\label{thm:one_multiple_cs}
		Customer surplus with LTE-U is nondecreasing as the amount of unlicensed bandwidth $W$ increases but is always less than the consumer surplus achieved without LTE-U. 
	\end{theorem}

	Theorem \ref{thm:one_multiple_cs} shows that adding more unlicensed spectrum benefits customers while adopting LTE-U always hurts customers when there are multiple entrant SPs in the market. This is because increasing the bandwidth of unlicensed spectrum actually increases the amount of resources for both the incumbent and entrant SPs.  However, when LTE-U is used, there is some loss in the total equivalent bandwidth and the incumbent is able to increase its price, leading to a loss in customer surplus.  
	

	So far we have seen that with multiple entrants, LTE-U increases firm profits (namely those of the incumbent) but decreases consumer welfare.  We next consider the impact of this technology on the overall welfare which includes both of these factors.  We first use a simplified example to gain insight  and then give a more general result. 
	
	Consider a homogeneous inverse demand function and linear congestion cost, i.e.,  $P(q) = T$ if $q\le A$, $P(q) = 0$, otherwise, and $g(x)=x$. Here, $A$ can be viewed as the size of the market while $T$ indicates the valuation of every consumer. We then have the following result.
	
	\begin{theorem}
		\label{thm:one_multiple_sw_homo}
		With a homogeneous inverse demand function and linear congestion,   if $W\le \frac{\sqrt{A^2+B^2T^2}-BT+A}{2T} $, social welfare will always increase when LTE-U is adopted for any $\alpha,\beta>0$.  Otherwise, social welfare can either increase or decrease when LTE-U is adopted.
	\end{theorem} 

 The proof is in Appendix \ref{app:one_multiple_sw_homo}. Theorem \ref{thm:one_multiple_sw_homo} shows that under the assumption of a homogeneous inverse demand function and linear congestion, when the unlicensed bandwidth $W$ is small, adopting LTE-U is good for social welfare no matter what $\alpha$ and $\beta$ are. Note that in these cases, customer surplus is always zero and all the social welfare comes from the revenue of the incumbent SP. When $W$ is beyond the threshold  in  the theorem, it becomes unclear how social welfare changes with LTE-U. It can depend on the choice of $\alpha$ and $\beta$.  Also note that the threshold bandwidth is an increasing function of $\frac{A}{T}$, which is the ratio between market size and customer valuation. When there are more customers in the market or the customer's valuation goes down,  the threshold goes up. This means that there is a larger range of $W$ for which LTE-U increases welfare.  This is because when there are more customers, or the customers have a lower valuation, the incumbent is incentivized to serve more customers to increase its revenue. 

The following results extend Theorem \ref{thm:one_multiple_sw_homo} to a  general inverse demand function and a general congestion function. 

	\begin{theorem}
		\label{thm:one_multiple_sw}
		Given a fixed $B>0$, $\alpha>0$ and $\beta>0$, There exists some $W_{th}>0$ such that when $W<W_{th}$ adopting LTE-U achieves a higher social welfare than that without LTE-U. But when $W$  is large enough, LTE-U always hurts social welfare.
	\end{theorem}

	A linear approximation method as in \cite{wang2017impact} is used to prove the theorem. When LTE-U is adopted, Theorem \ref{thm:one_multiple_cs} shows customer surplus decreases, which means the delivered price should increase. When the bandwidth of unlicensed spectrum is small, the increase in revenue of the incumbent is able to compensate for the customer surplus loss, so that the overall social welfare can increase. But when $W$ is large, the advantage of LTE-U may not be large enough to raise the delivered price to make up for the loss of customer surplus, which will result in a loss of social welfare. The detailed proof is in Appendix \ref{app:one_multiple_sw}.  
	
	\subsection{One incumbent \& one entrant}
	\label{sec:one_entrant}
	We next consider the case with only one entrant. If only this entrant is offering unlicensed service, then Lemma \ref{lemma:equ} no longer applies and so this case requires a separate analysis.  Before considering the impact of LTE-U, we first consider two possible ways the incumbent SP could act without this technology: it could compete with the entrant to serve customers on the unlicensed band or it could only serve customers on the licensed band.  We call the first case {\em unlicensed sharing} and in this case, the results are the same as when an incumbent without LTE-U competes with multiple entrants. We call the second case {\em licensed sharing}; in this case,   the entrant SP is able to use the unlicensed  spectrum exclusively. The objective of each SP is still to maximize revenue  while the Wardrop equilibrium conditions are satisfied. To be precise, in the licensed sharing case, the conditions for the entrant SP on the new band become
	\begin{eqnarray}
	& p_2+g\left(\frac{x_2}{W}\right)= P\left(x_1+x_2\right), {\rm if}\;\; x_2>0\nonumber\\
	\label{eqn:licensed_wardrop1}
	& p_2+g\left(\frac{x_2}{W}\right)\ge P\left(x_1+x_2\right),{\rm otherwise} .\nonumber
	\end{eqnarray}
	
	 We first give a brief result to compare the licensed sharing and unlicensed sharing cases without LTE-U.
	\begin{lemma}
		\label{lemma:exclusive_unlicense}
		In the case with one incumbent and one entrant SP, both the incumbent and entrant SPs are able to gain higher revenue with licensed sharing than with unlicensed sharing.
	\end{lemma}

	Lemma \ref{lemma:exclusive_unlicense} shows that rather than making the spectrum unlicensed, both the incumbent and entrant would prefer that it is exclusively licensed to the entrant SP. However, note that if the incumbent has the option of unlicensed sharing, then this will not be an equilibrium as it would always want to enter the unlicensed market and capture some of the entrant's revenue (even though this would eventually lead it to earning lower revenue).   
	
	Next we study of impact of LTE-U and in particular compare this to the licensed sharing case (which as noted above gives an upper bound on the incumbent's revenue in the unlicensed sharing case).  In this subsection, we assume a linear congestion function $g(x)=x$ and inverse demand function $P(x)=1-x$ to simplify the calculations and give some insights. 
	
	
	\begin{theorem}
		\label{thm:one_in_one_entrant_profit}
		With a linear congestion cost and inverse demand, we have the following comparisons with licensed sharing:
		\begin{enumerate}
			\item When $\frac{B}{1-\alpha}<\frac{4}{3}$, the incumbent SP can always gain higher revenue with LTE-U. Otherwise, the incumbent can be either better or worse off with LTE-U (depending on the parameter values).
			\item For any $\alpha,\beta\in(0,1)$, there always exists some $W_{th}$, such that when $W<W_{th}$, the incumbent SP can  gain higher revenue with LTE-U.
		\end{enumerate}		
	\end{theorem}

	The proof is in Appendix \ref{app:one_in_one_entrant_profit}. Both statements in  Theorem \ref{thm:one_in_one_entrant_profit}  give sufficient conditions to guarantee a larger revenue for the incumbent SP with LTE-U. Equations (\ref{eqn:Be}) and (\ref{eqn:We}) show that LTU-U  increases the amount resources of the incumbent and at the same time reduces the amount of resources of the entrant. Intuitively, this should lead to higher revenue for the incumbent with LTE-U.  The first statement in Theorem \ref{thm:one_in_one_entrant_profit} shows that this intuition holds when the incumbent's licensed spectrum is sufficiently small.  However, when there is a large enough amount  of licensed spectrum, the incumbent SP may suffer a loss in revenue with LTE-U. This is because the incumbent can already serve a large amount of customers on the licensed band and reducing the entrant's resources causes it to reduce the delivered price, lowering the incumbent's revenue. The second statement of  Theorem \ref{thm:one_in_one_entrant_profit} claims that as long as there is not too much unlicensed spectrum, the incumbent is always willing to adopt  LTE-U, which yields a higher revenue. That is because, when $W$ is relatively small, using LTE-U can decrease the equivalent bandwidth of the entrant competitor, which increases the congestion on unlicensed band significantly, giving  an advantage to the incumbent SP. But when $W$ is large, the decrease of the entrants' spectrum resource does not have a significant impact on congestion. As a result LTE-U can not increase the customer mass served by the incumbent enough to compensate for the lowered price due to competition. So the incumbent may not want to use LTE-U.

	\begin{figure}[htbp]
	\centering
	\includegraphics[scale=0.4]{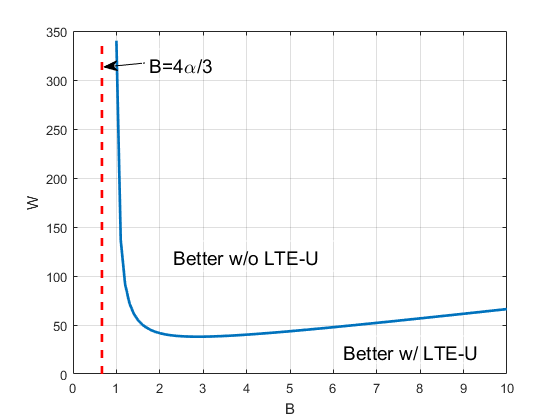}
	\caption{The bandwidth regions where LTE-U is better and worse for the incumbent's revenue with $\alpha = \frac{1}{2}$ and $\beta = \frac{1}{2}$.}
	\label{fig:range}
	\end{figure}

	We use Fig. \ref{fig:range} to illustrate the region where the incumbent SP can get more revenue with LTE-U. We choose $\alpha = \frac{1}{2}$ and $\beta = \frac{1}{2}$ in the figure. When  $B$ and $W$ lie below the blue curve, the incumbent is better off with LTE-U. The red dashed line represents $B= \frac{4\alpha}{3}$. The blue curve approaches to the red line asymptotically when $W\to \infty$. We can also see that the unlicensed bandwidth threshold $W_{th}$ is relatively large compared to the threshold for licensed bandwidth (the red dashed line).  That means in most practical cases, the incumbent SP would be willing to use LTE-U technology. 
	
	Note that when there is one entrant SP, using LTE-U can yield lower revenue for the incumbent in the licensed sharing case. However, in unlicensed sharing case, we have the following result.
	\begin{theorem}
		\label{thm:one_one_revenue_unlicensed}
		When there is only one entrant SP and unlicensed sharing is used, the incumbent always achieve higher revenue with LTE-U.
	\end{theorem}
	This result follows Theorem \ref{thm:one_multiple_profit} and Lemma \ref{lemma:exclusive_unlicense}. Detailed proof is in Appendix \ref{app:thm:one_one_revenue_unlicensed}. The reason is that when unlicensed sharing is used without LTE-U, the price on unlicensed band is zero due to the competition between the incumbent and entrant SP, which hurts the incumbent's revenue. If LTE-U is used, the incumbent does not compete with the entrant on unlicensed band directly. Instead, the incumbent accesses the unlicensed band through LTE-U. In this case, the service  price on unlicensed band is not  zero and the incumbent gains more revenue consequently. 
	
	Next we characterize the customer surplus in the case of one incumbent one entrant SP.
	\begin{theorem}
		\label{thm:cs_one_one}
		When there is one incumbent and one entrant SP, for any value of $B$, $\alpha$ and $\beta$, customer surplus is non-decreasing with unlicensed bandwidth $W$.
		Also, there exists some $W_{th}\ge B$, such that when $W\le W_{th}$, customer surplus decreases when LTE-U is adopted compared to licensed sharing and when  $W> W_{th}$, customer surplus increases.
	\end{theorem}

	The proof is in Appendix \ref{app:thm:cs_one_one}. The first result in Theorem \ref{thm:cs_one_one} is consistent with that in the multiple entrants case, which shows customer surplus increases with the bandwidth of unlicensed spectrum. The second result is  slightly different; it shows that LTE-U can hurt customer surplus when $W$ is relatively small but it is able to improve customer surplus when $W$ is large, while in the multiple entrants case, customer surplus always becomes worse with LTE-U. The reason is that when there is only one entrant SP, the price on the unlicensed band is not zero, which means a certain amount of consumers in the market are not served. When $W$ is large and LTE-U is used, the incumbent can use the additional unlicensed spectrum to alleviate congestion without hurting the entrant SP too much. As a result, more consumers in the market can be served and customer surplus increases.  Also note that when $W$ is relatively small, the loss in consumer surplus is balanced out by the increase in the SP's revenue. 
	
	However, when comparing to unlicensed sharing, the competition between the incumbent and entrant drives the price on unlicensed band to 0. As a result, the results in Theorem \ref{thm:one_multiple_cs} still holds if the incumbent and entrant SP use unlicensed sharing.

	We next examine how social welfare changes when LTE-U is adopted in the asymptotic case with $W\to\infty$.
	\begin{theorem}
		\label{thm:one_one_sw}
		In the case with one incumbent and one entrant SP, if $W\to\infty$, social welfare always increases when LTE-U is adopted.
	\end{theorem}

	The proof is in Appendix \ref{app:thm:one_one_sw}. Theorem \ref{thm:one_one_sw} shows that in the single entrant case, social welfare is higher with LTE-U when there is a large amount of unlicensed spectrum. Recall that in Theorem \ref{thm:one_multiple_sw}, we show in the case with multiple entrant SPs, social welfare is better with LTE-U when the bandwidth of unlicensed spectrum is small. The intuition is that   with multiple entrants,  the price is competed to 0 on the unlicensed band. As a result, the increment in revenue can only cover the loss of customer surplus when a small mass of customers are  served in the market, which implies a small $W$. However, in the single entrant case, the price is not zero, there can be a larger amount of customers unserved in the market, which leaves enough room for customer surplus to improve. When $W$ is large, the improvement of customer surplus can make up for the loss of revenue by the SPs.
	

	\section{Impact of different spectral efficiency}
	\label{sec:spectral_efficiency}
	
	Compared to 802.11ac, LTE-U offers a higher spectral efficiency due to features such as Hybrid ARQ and CSI feedback \cite{qualcomm,paolini2015lte}. In this section, we examine the market impact of these gains  in different  scenarios. 	 
	
	First, we look at the model used in the analysis. Without losing generality, we assume the spectral efficiency of WiFi is 1. We assume that the spectral efficiency of LTE is $\gamma \in [1,\infty)$.\footnote{We also include any MAC layer efficiency gains in this ``spectral efficiency" term.} A higher spectral efficiency leads to lower congestion on the same band, which we model by scaling the bandwidth on that band.  As a result, the congestion experienced by customers when using LTE-U is given by
	\begin{equation}
	\label{eqn:congestion_eff}
	\hat{g}_{in}{(x_1)}= \alpha g\left(\frac{ x_1}{\gamma\left(B+\beta W\right)}\right)+(1-\alpha) g\left(\frac{ x_1}{\gamma B}\right).
	\end{equation}
	And the resulting equivalent bandwidth of the incumbent SP is 
		\begin{equation}
	\label{eqn:Be_spec}
	B_e = \gamma \left[B+ \frac{\alpha\beta W}{1+\beta(1-\alpha)W/B}\right].
	\end{equation}
	The equivalent bandwidth of the entrant SP is still the same as (\ref{eqn:We}), because only LTE traffic is affected by $\gamma$.

	\subsection{Monopoly market}
	\label{sec:spectral_eff_mono}
	We again begin by looking at a monopoly market. In Section \ref{sec:mono}, we showed that the incumbent SP cannot gain more revenue with LTE-U. But when spectral efficiency is different between LTE and WiFi, this is no longer true as shown next: 
	\begin{theorem}
		\label{thm:mono_eff}
		In a monopoly scenario, the incumbent SP can gain more revenue by using the LTE-U if and only if 
		\begin{equation}
		\label{eqn:gamma_condition}
		\gamma> \frac{1+\beta(1-\alpha)W/B }{1-\beta(1-\alpha)}.
		\end{equation}
	\end{theorem}

	The proof is in Appendix \ref{app:thm:mono_eff}. This theorem shows that the gain in spectral efficiency of LTE-U over WiFi, $\gamma$, is large enough, then it is possible for a monopolist to gain more revenue with LTE-U. The reason is that when $\gamma$ is large, serving customers with LTE-U leads to less congestion, which allows the incumbent to serve more customers. We can also look at  equivalent bandwidth in (\ref{eqn:Be_spec}). When LTE-U is used, the increase of licensed equivalent bandwidth is multiplied by the spectral efficiency $\gamma$. When $\gamma$ is large enough, it is possible for the total equivalent bandwidth to increase, which helps the incumbent SP make more revenue. In this case, the incumbent SP is wiling to use LTE-U instead of serving customers on licensed and unlicensed bands separately.

	Note that when $\alpha$ and $\beta$ are fixed, the threshold of $\gamma$ in (\ref{eqn:gamma_condition}) increases with the ratio $\frac{W}{B}$. This means that when $W$ is relatively small comparing to $B$, the incumbent is willing to use LTE-U even if its advantage  over WiFi is not as large. However when $W$ is relatively large, the incumbent prefers serving customers on the two bands separately, because the gain from serving all customers with LTE cannot make up for the loss on the total  equivalent bandwidth. When the bandwidth of licensed and unlicensed spectrum is fixed, the threshold of  $\gamma$ in (\ref{eqn:gamma_condition}) is decreasing with $\alpha$. This shows that in a monopoly market, the incumbent is more likely to use LTE-U if the duty cycle is large. Moreover, the threshold of $\gamma$ is increasing with $\beta$. This means that the larger the portion of unlicensed spectrum the incumbent is allowed to use, the less profitable it will be  for the SP to use LTE-U. This is because if $\beta$ is large, the congestion on unlicensed band can be very high when the LTE-U mode is on, which causes a significant loss on both amount of customers served  and the revenue on the unlicensed band.  
	
	The next result characterized LTE-U's impact on customer surplus and total welfare. 
	
	\begin{theorem}
		\label{thm:mono_welfare_eff}
		In a monopoly scenario, both customer surplus and social welfare increase when LTE-U is used if and only if (\ref{eqn:gamma_condition}) holds. 
	\end{theorem}

	 The proof is in Appendix \ref{app:thm:mono_welfare_eff}. We see that the condition for customer surplus and social welfare to increase is the same as the condition for the incumbent's revenue to increase.  This is because  they all depend on the total equivalent bandwidth, which only increases when (\ref{eqn:gamma_condition}) holds. 
	 
	 Recall that in Section \ref{sec:mono}, customer surplus and social welfare both decrease if LTE-U is used when there is no difference on spectral efficiency between LTE and WiFi. However, if we assume LTE a better spectral efficiency than  WiFi, operating  LTE-U is able to increase the overall efficiency of spectrum usage and consequently can be beneficiary to customer surplus and social welfare.

	\subsection{Competition with multiple entrant SPs}
	\label{sec:multiple_ent_spectral}
	Now we turn to the case where multiple entrant SPs compete. We consider a linear model where the inverse demand function is $P(x) = 1-x$ and the congestion function is $g(x) = x$. 
	
	First we look at the incumbent's revenue with multiple entrant SPs. 
	\begin{theorem}
		\label{thm:one_multiple_profit_spectral}
		Consider one incumbent and multiple entrants. Given a fixed $\alpha >0$ and $\beta >0$, when LTE-U is adopted, for any $\gamma$, the  incumbent SP always gets a higher revenue and the customer mass served by entrant SPs decreases. Also the revenue of incumbent SP increases with  $\gamma$. 
	\end{theorem}

	The proof is in Appendix \ref{app:thm:one_multiple_profit_spectral}.As in Theorem \ref{thm:one_multiple_profit} (when $\gamma =1$),  LTE-U still results in an increase in the equivalent  bandwidth of the incumbent, which leads to an increase in its revenue.  For $\gamma >1$, this advantage only increases and grows with $\gamma$.

	While the incumbent's revenue behaves similarly for $\gamma >1$,  the result on customer surplus behaves differently when considering $\gamma >1$.  LTE-U basically operates as LTE on a certain portion of the unlicensed band. Thus, if LTE has a better spectral efficiency, it is possible to serve more customers in the market with the same amount of spectrum resources. As a result, customer surplus can increase in this case, which is different from the result in Theorem \ref{thm:one_multiple_cs}. We  characterize the customer surplus in the following theorem. 	
	
	\begin{theorem}
		\label{thm:one_multiple_cs_spectral}
		In the case with multiple entrants with a linear congestion cost and inverse demand function, for any $\alpha$ and $\beta$, if $B$ and $\gamma$ satisfy the following conditions:
		\begin{equation}
		\label{eqn:one_multiple_profit_spectral_condition1}
		B\le\frac{\left(\sqrt{2}-1\right)\left(1-\beta+ \alpha\beta\right)\alpha\beta}{2\left(1-\beta\right)},
		 \end{equation}
		 \begin{equation}
		 \label{eqn:one_multiple_profit_spectral_condition2}
		 \frac{\alpha\beta-2(1-K)B-\sqrt{\Delta}}{4(1-K)}\le\gamma\le \frac{\alpha\beta-2(1-K)B+\sqrt{\Delta}}{4(1-K)},
		\end{equation} 
		where $K=\frac{\alpha\beta}{1-\beta(1-\alpha)}$, and $\Delta = 2\alpha^2\beta^2-\left[2(1-K)B+\alpha\beta\right]^2$, there always exists some $W_{th}$, such that when $W<W_{th}$, customer surplus increases when LTE-U is used. Otherwise, customer surplus always decreases if LTE-U is used.
	\end{theorem}

	The proof is in Appendix \ref{app:thm:one_multiple_cs_spectral}. This theorem shows that when licensed bandwidth is small and LTE has a relatively large advantage over WiFi, using LTE-U can lead to serving  more customers. This also requires that the bandwidth of unlicensed spectrum is small. Because if $W$ is large, the loss on unlicensed band can no longer be compensated by the increase in the  efficiency. Note that the threshold for $\gamma$ in (\ref{eqn:one_multiple_profit_spectral_condition2}) shows that the efficiency gain cannot be too large, because when $\gamma$ is too large, the incumbent serves a large amount of customers even without LTE-U. There is no room for customer surplus to increase when LTE-U is used.

	For social welfare, the result in  Theorem \ref{thm:one_multiple_sw} still holds. A very simple way to verify this is that when $W$ is small, it is possible to increase customer surplus by using LTE-U. Since the incumbent's revenue increases with LTE-U and entrants always end up with no revenue, social welfare also increases when LTE-U is used. In the case with different spectral efficiency, it is possible to have customer surplus and social welfare increase at the same time, which is not possible when the spectral efficiency is the same.
	
	\subsection{Competition with one entrant SP}
	\label{sec:one_entrant_spectral_efficiency}
	We next consider the case with only one entrant. We focus on the licensed sharing case, where the entrant SP is able to use the unlicensed  spectrum exclusively unless the incumbent SP uses LTE-U. We first look at how the revenue of the incumbent changes when LTE-U is used.
	\begin{theorem}
		\label{thm:one_one_profit_spectral}
			With a linear congestion cost and inverse demand, when $\frac{\gamma B}{1-\alpha}<\frac{4}{3}$, the incumbent's  revenue is higher when LTE-U is used and is increasing with $\gamma$. 
	\end{theorem}

	The proof is in Appendix \ref{app:thm:one_one_profit_spectral}. Similar to Theorem \ref{thm:cs_one_one}, when $B$ is small, the incumbent is able to gain more revenue with LTE-U. But the threshold of $B$ decreases when we consider a spectral efficiency advantage of LTE. The intuition is again that when $\gamma >1$, , the equivalent bandwidth of the incumbent SP  increases.  When $\frac{4(1-\alpha)}{3\gamma}<B<\frac{4(1-\alpha)}{3}$, the increased equivalent bandwidth helps the incumbent to serve more customers to gain more revenue. In such a case,  it is possible that the increased equivalent bandwidth brought by LTE-U  reduces the revenue of incumbent, because it may lower the delivered price due to a more intense competition between the incumbent and entrant SPs. The theorem also states  that when the given condition holds, the incumbent's revenue increases with the spectral efficiency. However, when the given condition does not hold, a higher spectral efficiency may result in a loss of revenue.This is again caused by the increased  competition with the entrant SP.
	
	 Next we characterize the impact on customer surplus and social welfare.

	\begin{theorem}
		\label{thm:one_one_sw_cs_spectral}
		In the case with  one incumbent and one entrant SP, both  customer surplus and social welfare always increase with $\gamma$.
	\end{theorem}
	
	The proof is in Appendix \ref{app:thm:one_one_sw_cs_spectral}. Although it is possible for the incumbent to lose some revenue when $\gamma$ increases, customer surplus and social welfare always benefit from an increase in  spectral efficiency. This means that when there is a loss in revenue for the incumbent with increasing spectral efficiency, the gain in customer surplus is able to make up for the loss.
	
	\section{Impact of $\alpha$ and $\beta$}
	\label{sec:alpha}
	The duty cycle, $\alpha$, and the percentage of the band for LTE-U use, $\beta$, are two important parameters to maintain fair and efficient coexistence of LTE-U and other unlicensed spectrum users.  In this section, we investigate the impact of $\alpha$ and $\beta$ in both the cases with multiple entrants and with one entrant. To simplify our analysis we again assume no spectral efficiency gains (i.e., $\gamma =1$) and   again consider a linear model where the inverse demand function is $P(x) = 1-x$ and the congestion function is $g(x) = x$.

	\subsection{Impact of duty cycle}	
	First we consider when  $\beta$ is fixed and only vary the duty cycle $\alpha$ to see its impact. We begin with  the case of one incumbent and multiple entrant SPs in the market. Before proceeding with our analysis of varying the duty cycle, we give the following proposition which characterizes the market equilibrium in the assumed scenario.  
	\begin{proposition}
		\label{prop:ne}
		Assuming a  linear inverse demand function and congestion function and multiple entrants, the  equilibrium announced price of the incumbent SP and the customer mass served is 
		\begin{equation}
		p_1 = \frac{1}{2(1+W_e)},\;\;\; x_1 = \frac{B_e}{2(1+B_e+W_e)},\nonumber
		\end{equation}
		where $B_e$ and $W_e$ are defined in (\ref{eqn:Be}) and (\ref{eqn:We}) respectively. The announced prices of entrant SPs are all zero and the total customer mass served by the entrants is
		\begin{equation}
		w_t = \frac{W_e(2+2W_e+B_e)}{2(1+W_e)(1+B_e+W_e)}.\nonumber
		\end{equation} 
	\end{proposition}

	This proposition shows  that the equilibrium price can be expressed with the equivalent bandwidth in (\ref{eqn:Be}) and (\ref{eqn:We}).  We next investigate how the equivalent bandwidth $B_e$, $W_e$ and their sum change with the duty cycle $\alpha$. 
	\begin{lemma}
		\label{lemma:duty_cycle_band}
		The equivalent bandwidth $B_e$ increases with $\alpha$ and $W_e$ decreases with $\alpha$.
		If $W>\frac{B}{1-\beta}$, for $\alpha \in(0,\frac{1}{2})$, the total amount of equivalent bandwidth $B_e+W_e$ always decreases with $\alpha$.
	\end{lemma}


	The proof is in Appendix \ref{app:duty_cycle_band}. This lemma shows that when $B$ is relatively smaller than $W$, the total  equivalent bandwidth decreases with $\alpha$ in the range $(0,\frac{1}{2})$. As mentioned previously, the duty cycle is usually limited below $50\%$. That implies that in practice,  the total equivalent bandwidth decreases with $\alpha$. 
	\begin{theorem}
		\label{thm:dutycycle}
		When there is one incumbent SP and multiple entrant SPs, the revenue of the incumbent always increases with the duty cycle $\alpha$.
	\end{theorem}

	The proof is in Appendix \ref{app:thm:dutycycle}. Theorem \ref{thm:dutycycle} is a natural result of Lemma \ref{lemma:duty_cycle_band}. Because the incumbent SP gets more equivalent bandwidth with LTE-U while the entrants lose more resources with increasing $\alpha$, the incumbent's revenue  should increase with $\alpha$. Consequently, if there is no limit for choosing $\alpha$, the incumbent SP may want to raise $\alpha$ to a value close to 1.
	
	Things become different when we consider the case with only one entrant SP in the market. In this case, Lemma \ref{lemma:duty_cycle_band} still holds, but the   incumbent SP may not want to choose a large $\alpha$ all the time.  The following theorem describes such an example.
\begin{theorem}
	\label{thm:optimal_alpha}
	When there is only one incumbent and one entrant in the market and $W\to \infty$, the optimal $\alpha$ for the incumbent SP to maximize its revenue is $\alpha^* = \max\{1-\frac{3B}{4},0\}$. 
\end{theorem}

\begin{figure}[htbp]
		\centering
		\includegraphics[scale=0.4]{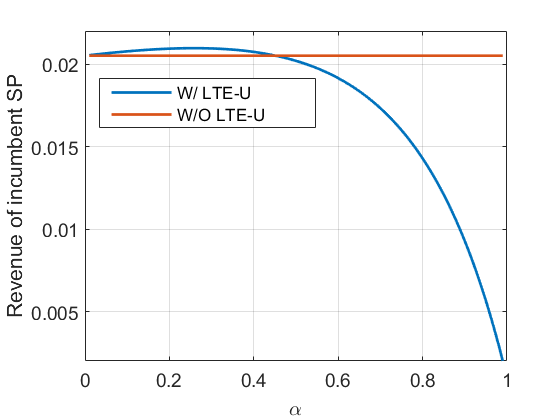}
		\caption{Revenue of the incumbent in the case with one entrant SP when $B=1$, $W\to\infty$ and $\beta =0.2$.}
		\label{fig:one_one_profit_alpha_B1_W1000}
	\end{figure}

	The proof is in Appendix \ref{app:thm:optimal_alpha}. Theorem \ref{thm:optimal_alpha} shows that the revenue of incumbent is no longer increasing with $\alpha$ when there is only one entrant SP in the market. Fig. \ref{fig:one_one_profit_alpha_B1_W1000} shows how revenue changes with $\alpha$ when $B = 1$, $W\to\infty$ and $\beta = 0.2$. We can see the revenue of the incumbent reaches a maximum when $\alpha = \frac{1}{4}$ and is higher than that without LTE-U. In the case with one entrant SP, the incumbent SP may want to choose a small $\alpha$ or  even does not want to use LTE-U technology ($\alpha^* = 0$) when there is plenty of licensed resource. Another thing to notice is that  when $W\to\infty$, the optimal $\alpha$ is non-increasing with licensed bandwidth $B$. This implies the more licensed spectrum the SP possesses, the smaller duty cycle it may prefer.  
	
	Next we look at the social welfare. Theorem \ref{thm:one_one_sw} states that in the case with one incumbent, one entrant and $W\to\infty$, LTE-U yields higher social welfare. We characterize the gap between the two cases in the following theorem. 
	\begin{theorem}
		\label{thm:sw_gap}
			When there is one incumbent and one entrant SP, if $W\to\infty$, the social welfare gap between the cases with and without LTE-U is non-decreasing in $\alpha$.
	\end{theorem}

	The proof is in Appendix \ref{app:thm:sw_gap}. Theorem \ref{thm:sw_gap} shows that when there is a sufficient amount  of unlicensed spectrum, a regulator may prefer a larger duty cycle $\alpha$, because it   increases the total amount of effective resources in the market, which results in a larger increase in social welfare.  
	
	\subsection{Fixed utilization ratio} 
	 In this section we fix the incumbent's utilization of the unlicensed spectrum, given by the product $\alpha\beta$.  We then study the impact of varying $\alpha$ and $\beta$ keeping this product fixed for the case of one incumbent and multiple entrants.  We set $\alpha\beta = k$, where $k$ is a constant. We then view  $\alpha$ as a variable in the analysis. In this case, $\alpha$ can vary in the range $(k,1)$. The equivalent bandwidths can be rewritten as 
	\begin{equation}
	\label{eqn:equivalant_bw}
	B_e = B+\frac{kW}{1+(k/\alpha-k)W/B},\; W_e = W-\frac{kW}{1-(k/\alpha-k)}.
	\end{equation}
	
	Note that both $B_e$ and $W_e$ are increasing in $\alpha$. As a result, the total amount of equivalent bandwidth increases with $\alpha$. But it remains unclear what impact this has on the incumbent's revenue.  The following theorem addresses this. 
	
	\begin{theorem}
		\label{thm:fix_uti_profit}
		In the case with one incumbent SP and multiple entrant SPs under the linear setting, if $\alpha\beta = k$ and $k$ is  a constant in $(0,1)$, then:
		\begin{enumerate}
			\item If $B>\frac{\sqrt{2}}{2}$ and $W\le B$, the incumbent's revenue  always decreases with $\alpha$  in the range $(k,1)$;
			\item For any choice $B$, there always exists some $W_{th}>0$ and $k_{th}\in(k,1)$, such that when $W>W_{th}$, the incumbent's revenue  decreases with $\beta$ in the range $(k,k_{th})$.
		\end{enumerate}
	\end{theorem} 

	The proof is in Appendix \ref{app:thm:fix_uti_profit}. Theorem \ref{thm:fix_uti_profit} shows that in different situations, the incumbent SP  has different preferences on higher $\alpha$ or $\beta$ when the product $\alpha\beta$ is fixed. When the bandwidth of the unlicensed spectrum is relatively small, the amount of equivalent bandwidth increases with $\alpha$, but the revenue decreases with $\alpha$. In this case, using a larger portion of unlicensed spectrum is more profitable than using a small portion for a longer time. However when $W$ is relatively large, the incumbent's revenue decreases with $\beta$ in some range, which implies that the incumbent may prefer a larger $\alpha$. In this case, a small portion of the spectrum may be enough for the incumbent to serve its customers. As a result a larger duty cycle $\alpha$ might be more profitable for the incumbent SP. 
	
\begin{figure}[htbp]
		\centering
		\subfigure[$B = 1$,$W = 1$ ]{
			\label{fig:one_multiple_fix_b1_W1_profit} 
			\includegraphics[width=1.68in]{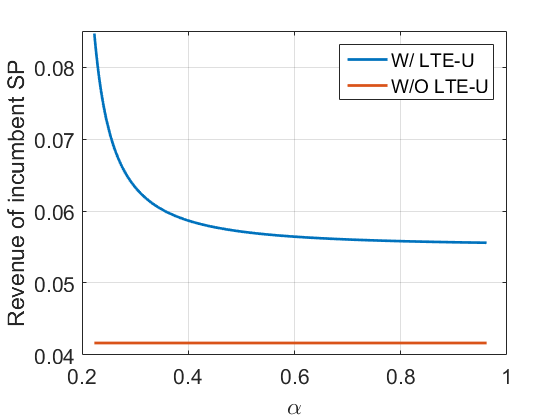}}
		\subfigure[$B = 1$,$W = 100$ ]{
			\label{fig:one_multiple_fix_b1_W100_profit} 
			\includegraphics[width=1.68in]{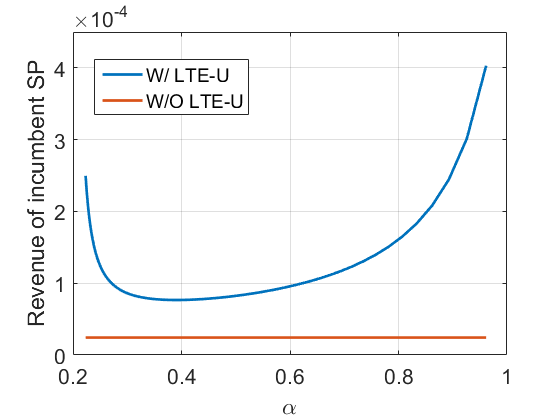}}
		\caption{Revenue of incumbent SP versus $\alpha$  with multiple entrant SPs and $k = 0.2$.}
		\label{fig:one_multiple_fix_profit} 
	\end{figure}

	Fig. \ref{fig:one_multiple_fix_profit} is an example of these two cases. We fix $k=0.2$ and $B=1$. We can see in Fig. \ref{fig:one_multiple_fix_b1_W1_profit}, when  $W$ is relatively small, the incumbent may prefer a lower $\alpha$. But when $W$ is relatively large, the incumbent may prefer a higher $\alpha$ as is shown in Fig. \ref{fig:one_multiple_fix_b1_W100_profit}.

	\begin{theorem}
		\label{thm:fix_uti_cs}
		In the case with one incumbent SP and multiple entrant SPs in the market, if $\alpha\beta = k$ and $k$ is a constant in $(0,1)$, then  customer surplus always increases with $\alpha$ in (k,1).
	\end{theorem} 
	
	From the expression of equivalent bandwidth in (\ref{eqn:equivalant_bw}), we know both $B_e$ and $W_e$ are increasing with $\alpha$, which implies the total amount of equivalent spectrum resources increases with $\alpha$. So in this case, a higher $\alpha$ can help increase the amount of virtual resources and  serve more customers. 


	\section{Numerical results}
	\label{sec:numerical}
	In this section we give some additional numerical examples illustrating our results. We again consider a  model with a linear inverse demand function and congestion function where $P(x) = 1-x$, $g(x) = x$. Both cases with fixed $\alpha,\beta$ and varying $\alpha, \beta$ are considered.
	
	\subsection{Fixed $\alpha$ and $\beta$}
\begin{figure}[htbp]
	\centering
	\subfigure[Revenue of incumbent ]{
		\label{fig:one_multiple_profit} 
		\includegraphics[width=1.68in]{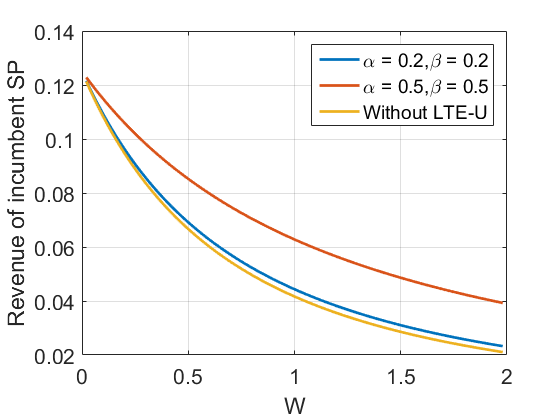}}
	\subfigure[Social welfare]{
		\label{fig:one_multiple_sw} 
		\includegraphics[width=1.68in]{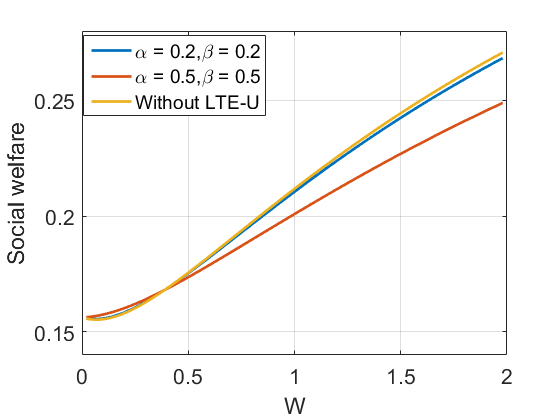}}
	\caption{Comparison of revenue and welfare versus $W$   in the case with multiple entrant SPs.}
	\label{fig:one_multiple_comparison} 
\end{figure}

	First we examine how the incumbent's revenue and  social welfare changes with the amount of unlicensed spectrum when there is  one incumbent SP and multiple entrant SPs in the market. We fix the licensed bandwidth as $B=1$ and set $\alpha,\beta$ to different values. The results are shown in Fig. \ref{fig:one_multiple_comparison}. As is described in Theorem \ref{thm:one_multiple_profit}, the incumbent is always gaining more revenue with LTE-U; this is illustrated  in Fig \ref{fig:one_multiple_profit}. Also we can see that when more spectrum can be used by LTE-U, and a higher duty cycle is allowed, the revenue is higher. The resulting social welfare is show in Fig. \ref{fig:one_multiple_sw}. We can see that when the bandwidth of additional unlicensed spectrum is small, social welfare increases slightly  with the adoption of LTE-U technology. But when more unlicensed spectrum is released, social welfare decreases with LTE-U. Another thing to notice is that social welfare decreases with $W$ when $W$ is small. This effect is also mentioned for the case without LTE-U  in \cite{nguyen2011impact}. The use of LTE-U makes  the social welfare loss smaller.  	
	
\begin{figure}[htbp]
		\centering
		\includegraphics[scale=0.4]{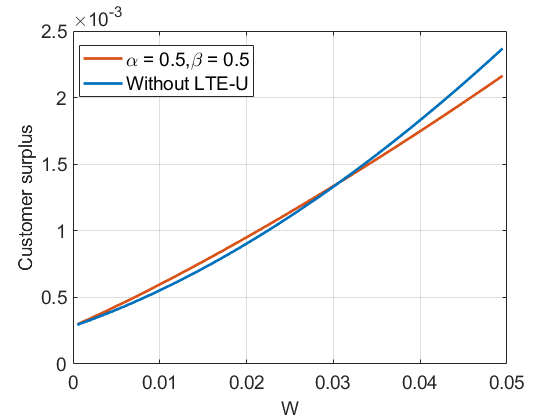}
		\caption{Customer surplus versus $W$  with $B=0.01$ and $\gamma = 5$}
		\label{fig:one_multiple_cs_spec}
	\end{figure}

	Next we look at the impact of different spectral efficiencies  when there are multiple entrant SPs. We showed that when spectral efficiency is not considered, customer surplus is always worse with LTE-U, but it is possible to end up with higher customer surplus when the efficiency difference is considered. In Fig. \ref{fig:one_multiple_cs_spec}, we show how customer surplus changes with $W$ while fixing $B$ and $\gamma$. We can see that when $W$ is below a certain value, customer surplus increases with LTE-U. However, for larger values of $W$, customer surplus is worse with LTE-U.
	
\begin{figure}[htbp]
		\centering
		\subfigure[Revenue]{
			\label{fig:one_one_profit_B5} 
			\includegraphics[width=1.68in]{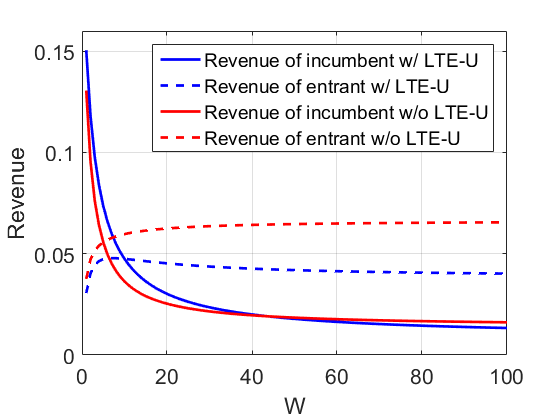}}
		\subfigure[Social welfare]{
			\label{fig:one_one_sw_B5} 
			\includegraphics[width=1.68in]{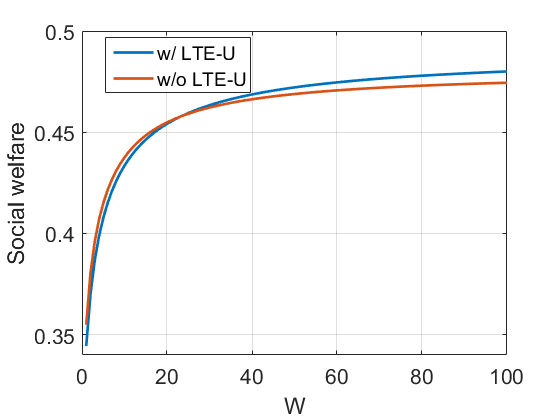}}
		\caption{Comparison of revenue and welfare versus $W$  in the case with one entrant SP.}
		\label{fig:one_one_B5} 
	\end{figure}

\begin{figure}[htbp]
		\centering
		\includegraphics[scale=0.4]{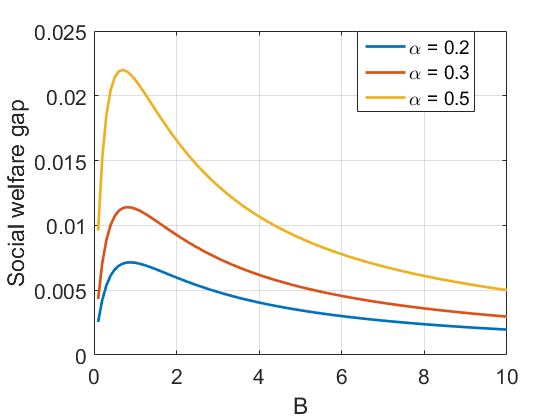}
		\caption{Social welfare gap between the cases with and without LTE-U as a function of $\alpha$ and $B$ for $W \rightarrow \infty$.}
		\label{fig:sw_gap}
	\end{figure}

	Next we take a look at the case with one incumbent and only one entrant SP in the market. We fix $B= 5$, $\alpha =0.5, \beta = 0.5$ and assume that $\gamma =1$ (i.e. there is not difference in spectral efficiency).  Results are shown in Fig. \ref{fig:one_one_B5}. We also include the entrant's revenue  in Fig. \ref{fig:one_one_profit_B5}. We can see that when $W$ is relatively small, the incumbent is able to gain more revenue with LTE-U while the entrant SP suffers a loss in revenue.  However, when $W$ is large, LTE-U hurts the revenue of both SPs.  The results for social welfare are shown in Fig. \ref{fig:one_one_sw_B5}. We can see when $W$ is large, social welfare increases with LTE-U and there is a social welfare gap between the cases with and without LTE-U. Next, we let $W\to\infty$ and see how this gap changes with $B$ under different $\alpha$ ($\beta$ makes no difference when $W\to\infty$). Results are shown in Fig. \ref{fig:sw_gap}. We can see that the social welfare gap first increases then decreases with $B$ and always increases with $\alpha$. When $B$ is small, LTE-U is able to increase the amount of spectrum resources of the incumbent SP to serve more customers, which benefits the incumbent's revenue  and customer surplus and as a result leads to a gain in  the social welfare. However, when $B$ is large, the gap is smaller  because both cases approach maximum possible social welfare in the market so that the increase in resources does not have as large an impact as when $B$ is smaller.  
	
	\subsection{Varying $\alpha$ and $\beta$}
\begin{figure}[htbp]
		\centering
		\subfigure[$B = 1$,$W = 0.1$ ]{
			\label{fig:one_multiple_alpha_b1_w01} 
			\includegraphics[width=1.68in]{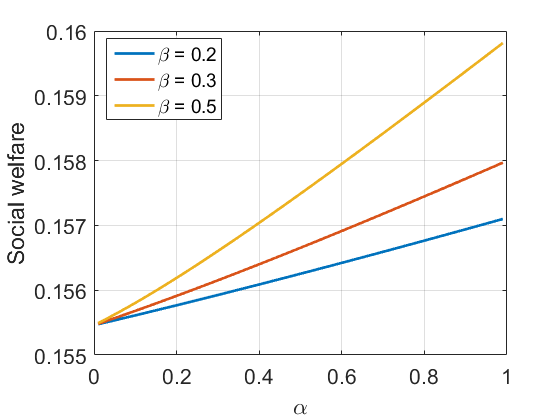}}
		\subfigure[$B = 1$,$W = 1$ ]{
			\label{fig:one_multiple_alpha_b1_w1} 
			\includegraphics[width=1.68in]{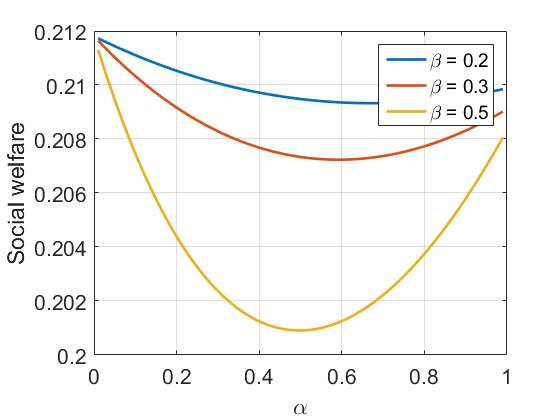}}
		\caption{Social welfare in the case with multiple entrants  with fixed $\beta$.}
		\label{fig:one_multiple_alpha} 
	\end{figure}

	 Next, we consider the impact of the duty cycle $\alpha$ with $\beta$ fixed when there are multiple entrants in the market. We have already shown that the incumbent's revenue  increases with $\alpha$ in Theorem \ref{thm:dutycycle}. In Fig. \ref{fig:one_multiple_alpha}, we  show how social welfare changes with $\alpha$ for different values of licensed bandwidth, unlicensed bandwidth, and $\beta$. We can see that when $W$ is small  (Fig. \ref{fig:one_multiple_alpha_b1_w01}), social welfare increase with the duty cycle $\alpha$.  In this case, a higher $\alpha$ is desirable by both the incumbent SP and a social planner. Also we can see that a larger $\beta$ helps increase the social welfare.  But when $W$ is slightly larger (Fig. \ref{fig:one_multiple_alpha_b1_w1}), social welfare first decreases then increases with $\alpha$. Additionally, we can see that when $\beta$ increases, social welfare decreases.

\begin{figure}[htbp]
	\centering
	\subfigure[$B = 1$,$W = 0.2$ ]{
		\label{fig:one_multiple_k_b1_w02} 
		\includegraphics[width=1.68in]{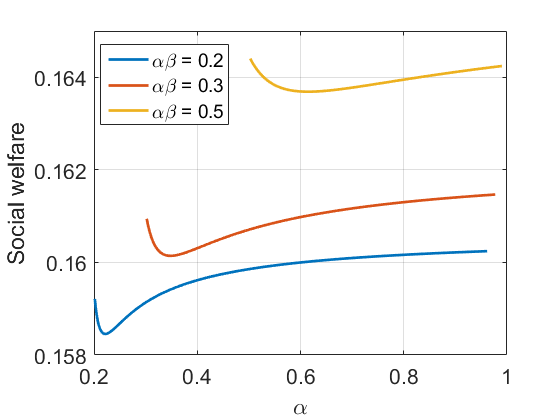}}
	\subfigure[$B = 1$,$W = 5$ ]{
		\label{fig:one_multiple_k_b1_w5} 
		\includegraphics[width=1.68in]{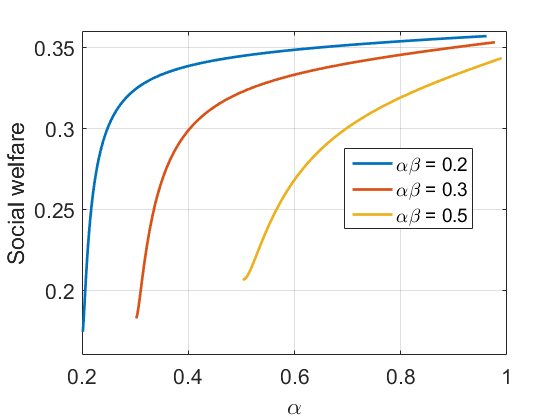}}
	\caption{Social welfare versus $\alpha$  in the case with multiple entrants and fixed $\alpha\beta$.}
	\label{fig:one_multiple_k} 
\end{figure}

	Next we show how social welfare changes with $\alpha$ when $\alpha\beta$ is fixed in the case with multiple entrant SPs.  Results are shown in Fig. \ref{fig:one_multiple_k}. We can see in Fig. \ref{fig:one_multiple_k_b1_w02} that when $W$ is relatively small, social welfare first decreases then increases with $\alpha$. That is because there is some welfare loss when adding a small amount of unlicensed spectrum to the market as is described in \cite{nguyen2011impact}. Recall that the equivalent bandwidth of unlicensed spectrum increases with $\alpha$ when $\alpha\beta$ is fixed. As a result the social welfare may suffer when $\alpha$ increases in the case of small $W$. But when $W$ is large as in Fig. \ref{fig:one_multiple_k_b1_w5}, social welfare always increases with $\alpha$.

	\section{Conclusion}	
	\label{sec:conclusion}
	In this paper, we analyzed the market impact of LTE-U technology on the competition among incumbent and entrant SPs with licensed and unlicensed spectrum. We first analyzed the case where the duty cycle, $\alpha$, and the portion of unlicensed spectrum that can be used by the incumbent, $\beta$, are fixed. Our results show that when there are multiple entrant SPs competing on the unlicensed band, the incumbent SP can get more revenue by using LTE-U. However when there is only one entrant SP in the market, the incumbent's revenue may decrease when LTE-U is adopted. We also show that the welfare impact of LTE-U depends on the market parameters - in some cases it can lead to a gain and in others a loss. We also investigated the case where LTE-U has a better spectral efficiency than WiFi.   In general, the incumbent benefits from the efficiency gain. When there are multiple entrants in the market both revenue and customer surplus can increase when LTE-U is used.  We also investigated  the impact of $\alpha$ and $\beta$ on the market. Our results show that when there are multiple entrants and if $\beta$ is fixed, the incumbent's revenue  increases with $\alpha$. However,  when there is only one entrant SP using unlicensed spectrum, the optimal $\alpha$ is not necessarily $1$ and can even be $0$. We also fixed the product $\alpha\beta$ to see whether the incumbent prefers a high $\alpha$ or a high $\beta$. Results show that when the unlicensed bandwidth is relatively small, the incumbent prefers high $\beta$ and when the unlicensed bandwidth is relatively large, the incumbent may prefer high $\alpha$.
	
	There are many ways this work could be extended. Extensions include considering the investment costs for a SP to upgrade to LTE-U, competition among multiple incumbents and different types of customers.  

	\bibliographystyle{IEEEbib}
	\bibliography{mybib}


	%
	%
	


	\appendices
	\section{Proof of Theorem \ref{thm:mono}}
	\label{append:thm:mono}
		\begin{proof}
			First, we claim that when congestion on both bands decreases, the optimal revenue of the SP increases; this is because  the SP can just announce the same price and attract more customers while keeping the potential to increase revenue by adjusting its price. As a result we only need to show that when the congestion level meets the lower bound, i.e., with linear congestion cost, the incumbent SP can gain no higher revenue than that without LTE-U.
			
			We use the equivalent licensed and unlicensed bandwidth to rewrite the optimization in (\ref{eqn:mono_optimization}).
			\begin{eqnarray}
			\label{eqn:mono_optimization1}
			\max_{p_1^l,p_1^u}&&p_1^lx_1^l +p_1^ux_1^u \nonumber\\
			{\rm s.t.} && p_1^l + g\left( \frac{x_1^l}{B_e}\right) = P(x_1^l+x_1^u),\nonumber\\
			&& p_1^u +   g\left( \frac{x_1^u}{W_e}\right) = P(x_1^l+x_1^u),\nonumber\\
			&&  p_1^l,p_1^u\ge0. \nonumber
			\end{eqnarray}
			For the optimization problem above, we can equivalently  use $x_1^l$ and $x_1^u$ instead of $p_1^l$ and $p_1^u$ as optimization variables. 
			
			
			From the first order conditions of the objective function over $x_1^l$ and $x_1^u$, we can show $\frac{x_1^l}{B_e}=\frac{x_1^u}{W_e}=\frac{x_1^l+x_1^u}{B_e+W_e}$, which means the congestion levels on the licensed and unlicensed bands are the same. Since we can verify $B_e+W_e\le B+W$, the total customer mass served becomes less when LTE-U  is applied. As a result,  the revenue of the incumbent SP decreases.  		
		\end{proof}

\section{Proof of Theorem \ref{thm:mono_welfare}}
\label{appendix:proof_mono_welfare}
In Appendix \ref{append:thm:mono}, we show that in the monopoly case, the congestion on the licensed and unlicensed bands are the same. As a result the price for service on these bands must also be the same. Thus, we can rewrite the problem in the following form:
\begin{eqnarray}
\label{eqn:mono_optimization_equivalent}
\max_{p}&&p_1 x_1 \\
{\rm s.t.} && p_1 + g\left( \frac{x_1}{B_e+W_e}\right) = P(x_1),\nonumber\\
&&  p_1,x_1\ge0. \nonumber
\end{eqnarray}
Again  using $x_1$ as the optimization variable, the  first order condition for optimality is 
\begin{eqnarray}
\label{eqn:first_order_proof_1}
\frac{\partial p_1x_1}{\partial x_1} &=&x_1P'(x_1)+P(x_1)- g\left( \frac{x_1}{B_e+W_e}\right)\nonumber\\
&&
- \frac{x_1}{B_e+W_e}g'\left( \frac{x_1}{B_e+W_e}\right)=0.
\end{eqnarray}
Since $P(x)$ is concave decreasing and $g(x)$ in convex increasing, the solution to (\ref{eqn:first_order_proof_1}) decreases when $B_e+W_e$ decreases. Again using that  $B_e+W_e<B+W$, we conclude that the total customer mass served decreases when LTE-U is used. 

Next we show customer surplus is an increasing function of the total customer mass served. Differentiating the  customer surplus  in (\ref{eqn:csdefinition}) with respect to the total customer mass served , $x$,  yields 
\begin{equation}
\label{eqn:d_cs1} 
\frac{\partial CS}{\partial x} =-x P'(x). \nonumber
\end{equation}
Again, because $P(x)$ is a concave decreasing function, customer surplus, $CS$, is always an increasing function in the total customer mass served. Therefore, customer surplus always decreases if LTE-U is used. 

Because both the incumbent's revenue and customer surplus decreases, the overall social welfare decreases as well.

	\section{Proof of Theorem \ref{thm:one_multiple_profit}}
	\label{appendix:proof_one_multiple_profit}
	\begin{proof}
		According to Lemma \ref{lemma:equ}, we have $p_i = 0$ for $i\ne 1$ at equilibrium. Thus the game can be reduced to the following optimization problem   
		\begin{eqnarray}
		\label{eqn:one_multiple_optimization1}
		\max_{p_1}&&p_1x_1 \\
		{\rm s.t.} && p_1 +  h_{in}(x_1) = P(x_1+w_t),\nonumber\\
		&& 0 +   h_{en}(w_t) = P(x_1+w_t),\nonumber\\
		&&  p_1\ge0, \nonumber
		\end{eqnarray}
		where $w_t$ is the total customer mass served on the unlicensed band and $h_{in}$ and $h_{en}$ denote the congestion associated with the incumbent and entrant SPs respectively. When LTE-U is adopted and $\alpha,\beta>0$, we have
		\begin{eqnarray}
		\label{eqn:congestion_eqv}
		h_{in}(x_1) = 	\alpha g\left(\frac{x_1}{B+\beta W}\right)+(1-\alpha) g\left(\frac{x_1}{B}\right)<g\left(\frac{x_1}{B}\right), \nonumber\\	    
		\label{eqn:h_en}
		h_{en}(w_t) =  \alpha g\left(\frac{w_t}{(1-\beta) W}\right)+(1-\alpha) g\left(\frac{w_t}{W}\right)> g\left(\frac{w_t}{W}\right).\nonumber
		\end{eqnarray}
		In addition, because $g(\cdot)$ is a convex increasing function, we have
		\begin{equation}
		\label{eqn:d_congestion}
		\frac{\partial h_{in}(x_1)}{\partial x_1} <\frac{\partial g\left(x_1/B\right)}{\partial x_1}.\nonumber
		\end{equation} 
		
		We then have the first order condition of revenue for the incumbent SP
		\begin{eqnarray}
		\label{eqn:foc_inc}
		\frac{\partial p_1x_1}{\partial x_1} &=&x_1P'(x_1+w_t) \frac{\partial (x_1+w_t)}{\partial x_1}\\	   
		&&P(x_1+w_t)- \left[h_{in}(x_1)+ \frac{\partial h_{in}(x_1)}{\partial x_1}\right].\nonumber
		\end{eqnarray}
		
		We use superscript $(\cdot)^{LTE-U}$ and ${(\cdot)}^{un}$ to denote the quantity in the case with and without LTE-U technology respectively.  First we use contradiction to show that $x_1^{LTE-U} >x_1^{un}$
		If we keep the selection of $x_1 = x_1^{un}$ or choose $x_1 < x_1^{un} $, base on the second equality constraint in (\ref{eqn:one_multiple_optimization1}) and $h_{en}(w_t)>g(w_t/W)$, we must have $w_t^{LTE-U}> w_t^{un}$. When $w_t$ decreases, the first order condition in (\ref{eqn:foc_inc}) becomes positive, because we have $h_{in}(x_1)<g(x_1/B)$ and $ \frac{\partial h_{in}(x_1)}{\partial x_1} <\frac{\partial g\left(x_1/B\right)}{\partial x_1}$.  It implies that the incumbent SP can still serve more customers to increase its revenue. Thus we have $x_1^{LTE-U} >x_1^{un}$, which contradicts with our assumption. As a result, we have $w_t^{LTE-U}< w_t^{un}$, because otherwise the second constraint in (\ref{eqn:one_multiple_optimization1}) will not hold, which proves the second statement.
		
		Additionally it can be verified that the total customer mass should satisfy $x_1^{LTE-U}+w_t^{LTE-U} < x_1^{un} +w_t^{un}$ so that the first order condition in (\ref{eqn:foc_inc}) can still be 0 when LTE-U is adopted. It proves the third statement. Given the conditions  $x_1^{LTE-U} >x_1^{un}$, $x_1^{LTE-U}+w_t^{LTE-U} < x_1^{un} +w_t^{un}$, the announced price $p_1$ must increase to guarantee that the first constraint in (\ref{eqn:one_multiple_optimization1}) still holds. Then the results on revenue in the first statement follows.	    
		
	\end{proof}
\section{Proof of Theorem \ref{thm:one_multiple_cs}}
	\label{app:thm:one_multiple_cs}
		\begin{proof}
		We take derivative of customer surplus  in (\ref{eqn:csdefinition}) with respect to the total customer mass served $Q$ and get
		\begin{equation}
		\label{eqn:d_cs} 
		\frac{\partial CS}{\partial Q} =-Q P'(Q). \nonumber
		\end{equation}
		Because $P(\cdot)$ is a concave decreasing function, customer surplus $CS$ is always an increasing function of $Q$. Theorem \ref{thm:one_multiple_profit}  shows that when the incumbent SP adopts the LTE-U technology, total customer mass decreases, which implies that customer surplus also decreases. 		
	\end{proof}

\section{Proof of Theorem \ref{thm:one_multiple_sw_homo}}
\label{app:one_multiple_sw_homo}
		\begin{proof}
		First  we consider the case without LTE-U. This case is analyzed in \cite{liu2014competitionopen}; we summarize the key results next: 
		\begin{enumerate}
			\item When $W\le\max\{\frac{A}{T}-\frac{B}{2},0\}$, customer surplus is zero and 
			\begin{small}
			\begin{equation}
			\label{eqn:sw_proof_1}
			SW = \frac{BT^2}{4};
			\end{equation}
			\end{small}
			\item When $\max\{\frac{A}{T}-\frac{B}{2},0\}<W\le\frac{\sqrt{A^2+B^2T^2}-BT+A}{2T}$, customer surplus is zero  and 
			\begin{small}
			\begin{equation}
			\label{eqn:sw_proof_2}
			SW = (A-WT)\left[T-\frac{(A-WT)}{B}\right];
			\end{equation}
		\end{small}
			\item When  $W>\frac{\sqrt{A^2+B^2T^2}-BT+A}{2T}$, we have positive customer surplus and 
			\begin{small}
			\begin{equation}
			\label{eqn:sw_proof_3}
			SW = AT-\frac{A^2(B+4W)}{4W(B+W)}.			
			\end{equation}
		\end{small}			
		\end{enumerate}	
		We next show that in the first two cases, the adoption of LTE-U increases the social welfare. Recall that with a linear congestion function, LTE-U increases $B$ to $B_e$ and decreases $W$ to $W_e$, where $B_e$ and $W_e$ are defined in (\ref{eqn:Be}) and (\ref{eqn:We}), respectively. 
		
		In the first case, when we change $B$ to $B_e$ and $W$ to $W_e$, since $W-W_e>B_e-B>0$, we always have $W_e\le\max\{\frac{A}{T}-\frac{B_e}{2},0\}$. This implies that $(B_e,W_e)$ still falls in the region of case 1), so that we can still use equation (\ref{eqn:sw_proof_1}) to calculate the social welfare. Obviously, when $B$ increases to $B_e$, social welfare also increases. 
		
		In the second case, we claim that when LTE-U is adopted, the equivalent unlicensed bandwidth $W_e$ can never go beyond the boundary $\frac{\sqrt{A^2+B_e^2T^2}-B_eT+A}{2T}$. Consider the following function :
		\begin{small}
		\begin{equation}
		f(b)= \frac{\sqrt{A^2+b^2T^2}-bT+A}{2T}.\nonumber
		\end{equation}
		\end{small}		 			
		 Given that $f(B) > W$ and $W-W_e>B_e-B$, we have 
		 \begin{small} 
		\begin{eqnarray}
		 \label{eqn:boundary}
		 &&f(B_e) = f(B)+\int_{B}^{B_e} f'(b)db\nonumber\\
		 &&> f(B) + \int_{B}^{B_e} -1 db= f(B) - (B_e-B) \nonumber\\
		 &&\ge W- (W-W_e) = W_e.\nonumber
		 \end{eqnarray}
		 \end{small}
		So in this case, when LTE-U is adopted, $B_e$ and $W_e$ can only fall into case 1) and case 2). Since all of the social welfare functions increase with $B$, it suffices to show that when reducing $W$ to $W_e$, social welfare is nondecreasing. When fixing $B$, it can be shown that social welfare in (\ref{eqn:sw_proof_2})  is decreasing in $W$ when $W\ge\frac{A}{T}-\frac{B_e}{2}$ and achieves maximum $\frac{BT^2}{4}$ when $W=\frac{A}{T}-\frac{B_e}{2}$. Consequently, when decreasing $W$ to $W_e$, if it still falls in the range of case 2), social welfare increases. If it falls into the range of case 1, it then becomes a constant with respect to unlicensed bandwidth $W$. As a result social welfare is nondecreasing when decreasing $W$ to $W_e$. 
		
	\end{proof}

\section{Proof of Theorem \ref{thm:one_multiple_sw}}
\label{app:one_multiple_sw}
	\begin{proof}		
	When $W\to 0$, social welfare is the same with and without LTE-U. In this limit, we  have $w_t\to0$ and $x_1\to x^*$, where $x^*$ represents the monopoly case optimal customer mass. Note that here $x^*$ is a constant if $B$ is fixed and can be characterized by the following equation
	\begin{equation}
	\label{eqn:monopolyoptimal}
	x^*P'(x^*)+P(x^*) = g\left(\frac{x^*}{B}\right)+\frac{x^*}{B}g'\left(\frac{x^*}{B}\right).
	\end{equation}
	Consequently, when $W\to 0$, we have $SW^{LTE-U}=SW^{un}$, where $SW^{LTE-U}$ and $SW^{un}$ denote the social welfare  with and without LTE-U, respectively. Hence, it is sufficient to show that $\lim\limits_{W\to0} \frac{\partial SW^{LTE-U}}{\partial W} >\lim\limits_{W\to0} \frac{\partial SW^{un}}{\partial W} $. Here, we use a linear approximation method as in \cite{wang2017impact} to characterize the  social welfare for both cases when $W\to0$.
	
	In both cases, differentiating the social welfare with respect to $W$  and letting $W\to0$ gives
	\begin{eqnarray}
	\label{eqn:pricesw}
	\lim\limits_{W\to0} \frac{\partial SW}{\partial W} = \left[-x^*P'(x^*)\right]\lim\limits_{W\to0}\frac{\partial x_1}{\partial W}.
	\end{eqnarray}
	
	Next we show how to calculate the value of $\lim\limits_{W\to0}\frac{\partial x_1}{\partial W}$. First, consider the case without LTE-U. In this case, the incumbent's revenue maximization problem can be transformed to the following optimization formulation
	\begin{eqnarray}
	\label{eqn:priceoptimization2}
	\max_{x_1}&&\left[P(x_1+\Delta w_t^{un})-g\left(\frac{x_1}{B}\right)\right]x_1,\\
	{\rm s.t.}&&x_1\ge0,\nonumber
	\end{eqnarray}
	where $\Delta w_t ^{un}= g^{-1}\left[P(x^*)\right]W$ is the customer mass increment on the unlicensed band  as $W \rightarrow 0$. Here $g^{-1}(\cdot)$ is the inverse function of $g(\cdot)$ and is well defined, because $g(\cdot)$ is an increasing function.

	The first order optimality condition for the optimization problem (\ref{eqn:priceoptimization2}) is
	\begin{small}
		\begin{eqnarray}
		\label{eqn:optimalcondition}
		&&(x^*+\Delta x_1^P)P'(x^*+\Delta x_1^P+\Delta w_t^P)+P(x^*+\Delta x_1^P+\Delta w_t^P)\nonumber \\
		&&= g\left(\frac{x^*+\Delta x_1^P}{B}\right)+\frac{x^*+\Delta x_1^P}{B}g'\left(\frac{x^*+\Delta x_1^P}{B}\right).
		\end{eqnarray}
	\end{small}
	Linearly approximating  each term in (\ref{eqn:optimalcondition}) at point $x^*$ and applying equation (\ref{eqn:monopolyoptimal}), we find:
	\begin{equation}
	\label{eqn:pricedeltax}
	\Delta x_1^{un} = \frac{\left[P'(x^*)+x^*P''(x^*)\right]g^{-1}\left[P(x^*)\right]W}{\left[\frac{2g'\left(\frac{x^*}{B}\right)}{B}+\frac{x^*g''\left(\frac{x^*}{B}\right)}{B^2}-2P'(x^*)-x^*P''(x^*)\right]}.\nonumber
	\end{equation}
	
	We can use similar method in the case with LTE-U technology. The resulting $\Delta x_1^{LTE-U}$ is 
	\begin{equation}
	\label{eqn:pricedeltax_lteu}
	\Delta x_1^{LTE-U} = \frac{\left[P'(x^*)+x^*P''(x^*)\right]h^{-1}\left[P(x^*)\right]W}{\left[\frac{2g'\left(\frac{x^*}{B}\right)}{B}+\frac{x^*g''\left(\frac{x^*}{B}\right)}{B^2}-2P'(x^*)-x^*P''(x^*)\right]},\nonumber
	\end{equation} 
	where $h(t)$ is defined as 
	\begin{equation}
	\label{eqn:ht}
	h(t) = (1-\alpha)g(t)+\alpha g\left(\frac{t}{1-\beta}\right).\nonumber
	\end{equation}
	It can be seen that $h(t)$ is also a convex increasing function and has a well defined inverse function $h^{-1}(t)$. Since we always have $h(t)>g(t)$ for $\alpha,\beta>0$, we have $g^{-1}\left[P(x^*)\right]>h^{-1}\left[P(x^*)\right]>0$. By substituting these values back to equation (\ref{eqn:pricesw}), we have
	\begin{eqnarray}
	\label{eqn:sw_comparison}
	\lim\limits_{W\to0} \frac{\partial SW^{LTE-U}}{\partial W}>\lim\limits_{W\to0} \frac{\partial SW^{un}}{\partial W}.\nonumber
	\end{eqnarray}		
	Therefore,  we can reach the conclusion that for some small $W$, social welfare increases when LTE-U is used.	
	
	Next we show that when $W$ is large enough, social welfare decreases when LTE-U is used. From the definition of the congestion function, we have $\lim\limits_{W\to \infty }g\left(\frac{w}{W}\right)=0$, because the customer mass is always bounded. When LTE-U is used, because both $\alpha$ and $\beta$ are fixed, the average congestion on the unlicensed band is
	\begin{equation}
	\label{eqn:zero_congestion}
	\lim\limits_{W\to \infty }\left[\alpha g\left(\frac{w}{(1-\beta) W}\right)+(1-\alpha) g\left(\frac{w}{W}\right)\right]	 = 0.\nonumber
	\end{equation}
	The Wardrop equilibrium	conditions suggests the delivered price is 0 in such a condition, which leads to 0 profit for the incumbent SP. Then, from the definition of social welfare, we have
	\begin{equation}
	\label{eqn:limit_sw}
	\lim\limits_{W\to\infty} SW =	\lim\limits_{W\to\infty}\sum\limits_j p_jx_j = CS. \nonumber
	\end{equation}
	Theorem \ref{thm:one_multiple_cs} states that customer surplus always decreases when LTE-U is used. Hence, social welfare also decreases in such a case. 
\end{proof}

\section{Proof of Theorem \ref{thm:one_in_one_entrant_profit}}
\label{app:one_in_one_entrant_profit}
	\begin{proof}
		We first give the equilibrium of the incumbent SP and entrant SP in terms of the equivalent bandwidth. 
		\begin{eqnarray}
		\label{eqn:one_one_equilibrium}
		p_1 = \frac{2+2B_e+W_e}{4+4B_e+4W_e+3B_eW_e},x_1 = \frac{B_e(1+W_e)}{1+B_e+W_e}p_1;\\
		p_2 = \frac{2+B_e+2W_e}{4+4B_e+4W_e+3B_eW_e},x_2 = \frac{W_e(1+B_e)}{1+B_e+W_e}p_2.\nonumber
		\end{eqnarray}
		When LTE-U is adopted, the spectrum resources of SPs changes from $B$ and $W$ to $B_e$ and $W_e$ respectively. It can be verified that $\frac{\partial p_1x_1}{\partial W_e}
		\le 0$ for any pair of $B_e$ and $W_e$. Since we have $W_e<W$, changing from $W$ to $W_e$ always increases the revenue of incumbent. 
		
		The derivative of revenue with respect to $B_e$ can be written in the following form 
		\begin{equation}
		\label{eqn:diff_Be}
		\frac{\partial p_1x_1}{\partial B_e} = T_0\left[M_0+(28 + 10 B_e - 6 B_e^2) W_e^4 + (4 - 3 B_e) W_e^5\right], \nonumber
		\end{equation}
		where $T_0,M_0>0$. It can be verified that if $4-3B_e>0$, $\frac{\partial p_1x_1}{\partial B_e}>0$ always holds. From (\ref{eqn:Be}), we know $B_e\in(B,\frac{B}{1-\alpha})$ for any $\alpha,\beta\in(0,1)$. As a result,  $\frac{B}{1-\alpha}<\frac{4}{3}$ is a sufficient condition for $\frac{\partial p_1x_1}{\partial B_e}>0$. Because adopting LTE-U increases $B$ to $B_e$, it implies that the incumbent's revenue increases when $\frac{B}{1-\alpha}<\frac{4}{3}$ holds. 
		
		To prove the second statement, we transform the derivative to another form.
		\begin{eqnarray}
		\frac{\partial p_1x_1}{\partial B_e} =T_1 \left[M_1+B_e^2 (48 + 88 W_e + 32 W_e^2 - 6 W_e^3)\right. \nonumber\\
		\label{eqn:diff_Be2}
		+ \left.B_e (48 + 116 W_e + 84 W_e^2 + 13 W_e^3 - 3 W_e^4)\right],\nonumber
		\end{eqnarray}
		where $T_1,M_1>0$. Since $W_e<W$, it is straightforward that when $W$ is small enough to satisfy $48 + 88 W + 32 W^2 - 6 W^3>0$  and $48 + 116 W + 84 W^2 + 13 W^3 - 3 W^4>0$, we always have $\frac{\partial p_1x_1}{\partial B_e}> 0$. As a result the revenue of incumbent increases with LTE-U in this case.
		
	\end{proof}

\section{Proof of Theorem \ref{thm:one_one_revenue_unlicensed}}
\label{app:thm:one_one_revenue_unlicensed}
\begin{proof}
	We use Theorem \ref{thm:one_multiple_profit} and Lemma \ref{lemma:exclusive_unlicense} to prove the result. First, if the incumbent is allowed to use LTE-U and also serve customers on unlicensed band, the result follows Theorem \ref{thm:one_multiple_profit}, because both the incumbent and entrant SP are competing on the unlicensed band, which drives the price to zero. In this case using LTE-U yields higher revenue according to the result in Theorem \ref{thm:one_multiple_profit}. Next, from Lemma \ref{lemma:exclusive_unlicense}, we know that the incumbent gains higher revenue in the licensed sharing case than unlicensed sharing with equivalent bandwidth $B_e$ and $W_e$. Consequently, we conclude that the incumbent can gain higher revenue with LTE-U than without LTE-U in the one entrant unlicensed sharing case.
\end{proof}

\section{Proof of Theorem \ref{thm:cs_one_one}}
\label{app:thm:cs_one_one}
	\begin{proof}		
	In Appendix \ref{app:one_in_one_entrant_profit}, we give the equilibrium price and customer mass served by the incumbent and entrant SPs in (\ref{eqn:one_one_equilibrium}). In this proof, we can still use those results. First we show that customer surplus is non-decreasing with $W$. As is proved in Appendix \ref{app:thm:one_multiple_cs}, we only need to look at the total customer mass served by SPs in the market, because customer surplus is an increasing function on total customer mass served. We have
	\begin{eqnarray}
	\label{eqn:customer_mass_diff}
	\frac{\partial (x_1+x_2)}{\partial W}=\frac{\partial (x_1+x_2)}{\partial W_e}\frac{\partial W_e}{\partial W}+ \frac{\partial (x_1+x_2)}{\partial B_e}\frac{\partial B_e}{\partial W}.\nonumber
	\end{eqnarray}
	From (\ref{eqn:Be}) and (\ref{eqn:We}), we have
	\begin{eqnarray}
	\label{eqn:d_Be_We}
	\frac{\partial B_e}{\partial W} &=&\frac{ (\alpha\beta B^2)}{(B + (1 - \alpha) \beta W)^2}\ge 0,\nonumber\\
	\frac{\partial W_e}{\partial W}&=& \frac{1-\beta}{1-\beta(1-\alpha)} \ge 0.\nonumber		
	\end{eqnarray}		
	From (\ref{eqn:one_one_equilibrium}), we have 
	\begin{footnotesize}
		\begin{eqnarray}
		\label{eqn:d_x1x2}
		\frac{\partial (x_1+x_2)}{\partial B_e} &=&\frac{2 (1 + W_e)^2 (4 + 4 W_e+ 3 W_e^2)}{(1 + B_e + W_e)^2 (4 + 4 B_e + 
			4 W_e + 3 B_e W_e)^2}\nonumber\\
		&&+\frac{ B_e^2 (8 + 16 W_e + 9 W_e^2)}{(1 + B_e + W_e)^2 (4 + 4 B_e + 
			4 W_e + 3 B_e W_e)^2}\nonumber\\
		&&+\frac{
			4 B_e (4 + 10 W_e + 9 W_e^2 + 3 W_e^3)}{(1 + B_e + W_e)^2 (4 + 4 B_e + 
			4 W_e + 3 B_e W_e)^2}\nonumber\\
		&\ge& 0,\nonumber\\
		\frac{\partial (x_1+x_2)}{\partial W_e} 
		&=&\frac{B_e^2 (30 + 36 W_e + 9 W_e^2)}{(1 + B_e + W_e)^2 (4 + 4 B_e + 
			4 W_e + 3 B_e W_e)^2}\nonumber\\
		&&+\frac{ 8 B_e (3 + 5 W_e + 2 W_e^2)}{(1 + B_e + W_e)^2 (4 + 4 B_e + 
			4 W_e + 3 B_e W_e)^2}\nonumber\\
		&&+\frac{6 B_e^4 + 8 (1 + W_e)^2 + 4 B_e^3 (5 + 3 W_e)}{(1 + B_e + W_e)^2 (4 + 4 B_e + 
			4 W_e + 3 B_e W_e)^2}\nonumber\\
		&\ge& 0.\nonumber	
		\end{eqnarray}
	\end{footnotesize}
	As a result we conclude $\frac{\partial (x_1+x_2)}{\partial W}\ge 0$  and consequently, customer surplus is non-decreasing with $W$.
	
	Next, we prove the second statement. We use $x_{total}^{LTE-U}$ and $x_{total}^{un}$ to denote the total customer mass served with and without LTE-U, respectively. We have 
	\begin{eqnarray}
	\label{eqn:customer_mass_difference1}
	\lim\limits_{W\to0} (x_{total}^{LTE-U} - x_{total}^{un}) &=&0,\nonumber \\
	\lim\limits_{W\to \infty}(x_{total}^{LTE-U} - x_{total}^{un}) &=&\frac{2+3B_e}{4+3B_e}-\frac{2+3B}{4+3B} >0.\nonumber 
	\end{eqnarray}
	Also we have	
	\begin{equation}
	\label{eqn:customer_mass_difference}
	\lim\limits_{W\to 0}\frac{\partial (x_{total}^{LTE-U} - x_{total}^{un})}{\partial W} < 0.\nonumber 
	\end{equation}
	Additionally, it can be verified that there is a unique $W>0$ such that $\frac{\partial (x_{total}^{LTE-U} - x_{total}^{un})}{\partial W}=0$.  Therefore, we can find some threshold, such that when $W$ is below the threshold,  $\frac{\partial (x_{total}^{LTE-U} - x_{total}^{un})}{\partial W}$ is negative and when $W$ is above the threshold, $\frac{\partial (x_{total}^{LTE-U} - x_{total}^{un})}{\partial W}$ is positive. Note that  $\lim\limits_{W\to0} (x_{total}^{LTE-U} - x_{total}^{un}) =0$ and $\lim\limits_{W\to \infty}(x_{total}^{LTE-U} - x_{total}^{un}) >0$. Also all terms in $x_{total}^{LTE-U} - x_{total}^{un}$ is continuous in $W$. Consequently, there exists $W_{th}$, such that when $W<W_{th}$, $x_{total}^{LTE-U} - x_{total}^{un} < 0$ and when $W>W_{th}$, $x_{total}^{LTE-U} - x_{total}^{un} > 0$. Furthermore, we have 
	\begin{equation}
	\label{eqn:diff_bw}
	(x_{total}^{LTE-U} - x_{total}^{un})|_{W=B} <0 .\nonumber
	\end{equation}
	So we can conclude $W_{th}> B$.		
\end{proof}

\section{Proof of Theorem \ref{thm:one_one_sw}}
	\label{app:thm:one_one_sw}
	\begin{proof}
		When $W\to\infty$, we can find $\lim\limits_{W\to\infty}\frac{\partial SW}{\partial B} =\frac{8+3B}{(4+3B)^2}\ge0 $ and $\lim\limits_{W\to\infty}B_e = \frac{B}{1-\alpha}$. Adopting LTE-U expands $B$ to $B_e$. Thus social welfare increases when LTE-U is adopted. 
	\end{proof}

\section{Proof of Theorem \ref{thm:mono_eff}}
	\label{app:thm:mono_eff}
		\begin{proof}  
		In Appendix \ref{append:thm:mono}, we show that in the monopoly case, the congestion on the licensed and unlicensed bands are the same. The revenue optimization problem is then in the form of (\ref{eqn:mono_optimization_equivalent}). Next, we show that the revenue of the incumbent always increases with the total equivalent bandwidth, $B_e+W_e$. When the total equivalent bandwidth increases, the incumbent can announce the same price and serve more customers, which yields higher revenue. The solution to the optimization problem only makes the revenue even higher.  So the revenue of the incumbent always increases with total equivalent bandwidth. Without LTE-U, the equivalent bandwidth is $\gamma B+W$. With LTE-U, the equivalent bandwidth is $B_e+W_e$, where $B_e$ is in (\ref{eqn:Be_spec}) and $W_e$ is in (\ref{eqn:We}). 	Then by solving the inequality 
		\begin{small}
			\begin{eqnarray}
			\gamma \left[B+ \frac{\alpha\beta W}{1+\beta(1-\alpha)W/B}\right]+ W -\frac{\alpha\beta W}{1-\beta(1-\alpha)}  
			> \gamma B+W\nonumber,
			\end{eqnarray}
	\end{small}	
	we get the condition that the incumbent's revenue increases, which is (\ref{eqn:gamma_condition}). 		
	\end{proof}

\section{Proof of Theorem \ref{thm:mono_welfare_eff}}
	\label{app:thm:mono_welfare_eff}
	\begin{proof}
	In Appendix \ref{appendix:proof_mono_welfare}, we show that in the monopoly case, customer surplus increases with the total equivalent bandwidth. In Appendix \ref{app:thm:mono_eff}, we show that the incumbent's revenue increases with total equivalent bandwidth. Because social welfare is the sum of incumbent's revenue and customer surplus, social welfare also increases with the total equivalent bandwidth. 
	
	In Appendix \ref{app:thm:mono_eff}, we also show that the total equivalent bandwidth increases if and only if  (\ref{eqn:gamma_condition}) holds. Therefore, we come to the conclusion.
\end{proof}

\section{Proof of Theorem \ref{thm:one_multiple_profit_spectral}}
\label{app:thm:one_multiple_profit_spectral}
	\begin{proof}		
	For the first statement, the proof follows the same steps as the proof of Theorem \ref{thm:one_multiple_profit} in Appendix \ref{appendix:proof_one_multiple_profit}. Those arguments still hold because for any $\gamma \ge 1$, the equivalent licensed bandwidth increases and the equivalent unlicensed bandwidth decreases when LTE-U is used. 
	
	The second statement can be proved by finding the equilibrium revenue assuming a linear inverse demand function and congestion function.  Since all entrant SPs announce zero price in equilibrium, the optimization problem for the incumbent SP can be written as
	\begin{eqnarray}
	\label{eqn:one_multiple_optimization_spec}
	\max_{p_1}&&p_1x_1 \\
	{\rm s.t.} && p_1 +  \frac{x_1}{B_e} = 1-(x_1+w_t),\nonumber\\
	&& 0 +   \frac{w_t}{W_e} = 1-(x_1+w_t),\nonumber\\
	&&  p_1\ge0. \nonumber
	\end{eqnarray}
	The resulting equilibrium price and customer mass is 
	\begin{equation}
	p_1 = \frac{1}{2(1+W_e)},\;\;\; x_1 = \frac{B_e}{2(1+B_e+W_e)}.\nonumber
	\end{equation}		
	It can be seen that $p_1$ is irrelevant to $B_e$ and $\frac{\partial{x_1}}{\partial B_e}>0$. Hence the incumbent's revenue increases with $B_e$. Because $B_e$ in (\ref{eqn:Be_spec}) also increases with $\gamma$. We conclude that  the revenue of incumbent SP increases with  $\gamma$.
	
\end{proof}

\section{Proof of Theorem \ref{thm:one_multiple_cs_spectral}}
\label{app:thm:one_multiple_cs_spectral}
	\begin{proof}
	To characterize the behavior of customer surplus, we look at the total customer mass served in the market. Because it is proved in Appendix \ref{app:thm:one_multiple_cs} that customer surplus is an increasing function in the total customer mass, we only need to look at the total customer mass.
	
	We use $x_{total}^{LTE-U}$ and $x_{total}^{un}$ to denote the total customer mass served with and without LTE-U, respectively. By solving the optimization problem in \ref{eqn:one_multiple_optimization_spec}, we can find the total customer mass 
	\begin{eqnarray}
	\label{eqn:customer_mass_total1}
	x_{total}^{un}  = 1- \frac{1}{2(1+W)}-\frac{1}{2(1+\gamma B+W)},\\
	\label{eqn:customer_mass_total2}
	x_{total}^{LTE-U} =  1- \frac{1}{2(1+W_e)}-\frac{1}{2(1+ B_e+W_e)},
	\end{eqnarray}
	where $B_e$ and $W_e$ are defined in (\ref{eqn:Be_spec}) and (\ref{eqn:We}), respectively.
	
	We look at the difference between  $x_{total}^{LTE-U}$ and $x_{total}^{un}$. It can be seen that 
	\begin{equation}
	\label{eqn:lim_diff}
	\lim \limits_{W\to\infty} (x_{total}^{LTE-U} -x_{total}^{un}) = 0.
	\end{equation}
	Also we can write the derivative of the difference in the following form:
	\begin{small}
		\begin{eqnarray}
		\label{eqn:2_order_diff}
		\frac{\partial ( x_{total}^{LTE-U} -x_{total}^{un})}{\partial  W}  &=& -\frac{1}{2(1+W)^2}-\frac{1}{2(1+\gamma B+W)^2}\\
		&& +\frac{K}{2(1+KW)^2}+\frac{K}{2(1+\gamma B+KW)^2} \nonumber,
		\end{eqnarray}
	\end{small}	
	where $K=\frac{\alpha\beta}{1-\beta(1-\alpha)}<1$ and is irrelevant to $W$. It can be seen from (\ref{eqn:2_order_diff}) that $\lim\limits_{W\to 0 } 	\frac{\partial ( x_{total}^{LTE-U} -x_{total}^{un})}{\partial  W}<0$ and $\lim\limits_{W\to \infty } 	\frac{\partial ( x_{total}^{LTE-U} -x_{total}^{un})}{\partial  W}=0$. 
	
	Additionally we need to examine when the derivative in (\ref{eqn:2_order_diff}) reaches 0. We can rewrite it in the following form:
	\begin{scriptsize}
		\begin{eqnarray}		
		\frac{\partial ( x_{total}^{LTE-U} -x_{total}^{un})}{\partial  W}  &=& \left[\frac{1}{2(\frac{1}{\sqrt{K}}+\sqrt{K}W)^2}-\frac{1}{2(1+W)^2}\right]  \nonumber\\
		&& +\left[\frac{1}{2(\frac{1+\gamma B}{\sqrt{K}}+\sqrt{K}W)^2}-\frac{1}{2(1+\gamma B+W)^2}\right] \nonumber\\
		\label{eqn:2_order_diff1}
		&\buildrel \Delta \over =& A_1 + A_2.
		\end{eqnarray}
	\end{scriptsize}
	Note that when $W\le \frac{1}{\sqrt{K}}$, both $A_1$ and $A_2$ in (\ref{eqn:2_order_diff1}) are negative and when $W\ge\frac{1+\gamma B}{\sqrt{K}} $, both $A_1$ and $A_2$ are positive. Since $A_1$ and $A_2$ are continuous functions of $W$, there must exist some $W\in(\frac{1}{\sqrt{K}}, \frac{1+\gamma B}{\sqrt{K}})$ to make (\ref{eqn:2_order_diff1}) zero. Also note that $A_1$ is increasing with $W$ and $A_2$ is decreasing with $W$, when $W\in(\frac{1}{\sqrt{K}}, \frac{1+\gamma B}{\sqrt{K}})$. The $W$ to make (\ref{eqn:2_order_diff1}) zero is unique.

	 Thus, $\frac{\partial  ( x_{total}^{LTE-U} -x_{total}^{un})}{\partial  W}$ is smaller than 0 when $W$ is below a certain threshold value and greater than 0 when $W$ is above the threshold value. Therefore, the difference $( x_{total}^{LTE-U} -x_{total}^{un})$ first decreases when $W$ is below the threshold value and then increases when $W$ is above the threshold value. 	Combining with  (\ref{eqn:lim_diff}), we can argue that if $W$ is above the threshold value, we have $x_{total}^{LTE-U} -x_{total}^{un} \le 0$. And if $\lim \limits_{W\to 0} (x_{total}^{LTE-U} -x_{total}^{un} )\le 0$, we have $x_{total}^{LTE-U} -x_{total}^{un} \le 0$ for all $W>0$. Otherwise we can find some $W_{th}$ such that when $W<W_{th}$, $(x_{total}^{LTE-U} -x_{total}^{un})>0$.

	Next, we need to find the conditions for $\lim \limits_{W\to 0} (x_{total}^{LTE-U} -x_{total}^{un})>0$. We have 
	\begin{eqnarray}
	\label{eqn:diff_0}
	&&\lim\limits_{W\to 0} (x_{total}^{LTE-U} -x_{total}^{un})\\& =& -2(1-K)\gamma^2 +\left[\alpha \beta -2(1-K)B \right]\gamma - (1-K)B^2, \nonumber
	\end{eqnarray}
	where  $K=\frac{\alpha\beta}{1-\beta(1-\alpha)}$. We view (\ref{eqn:diff_0}) as a quadratic function of $\gamma$. The discriminant is given by 
	\begin{equation}
	\label{eqn:discriminant}
	\Delta = 2\alpha^2\beta^2-\left[2(1-K)B+\alpha\beta\right]^2.
	\end{equation}
	To guarantee there exists real value $\gamma$ to make (\ref{eqn:diff_0}) greater than 0, we need $\Delta\ge0$, which gives us the condition in (\ref{eqn:one_multiple_profit_spectral_condition1}). Solving the inequality $\lim\limits_{W\to 0} (x_{total}^{LTE-U} -x_{total}^{un})\ge 0$ gives us the condition in (\ref{eqn:one_multiple_profit_spectral_condition2}). That completes the proof.	
\end{proof}

\section{Proof of Theorem \ref{thm:one_one_profit_spectral}}
\label{app:thm:one_one_profit_spectral}
	\begin{proof}
	In the case linear congestion cost and inverse demand, we can find the equilibrium price and customer mass served in the market. Following the steps in Appendix \ref{app:one_in_one_entrant_profit}, we know that if $4-3B_e>0$, we have $\frac{\partial p_1x_1}{\partial B_e} > 0$ and consequently, the incumbent's revenue always increases when LTE-U is used. When $\gamma$ is considered, we have $B_e\in(\gamma B,\frac{\gamma B}{1-\alpha})$ for any $\alpha, \beta \in(0,1)$. As a result, $\frac{\gamma B}{1-\alpha}<\frac{4}{3}$ is a sufficient condition for the incumbent's revenue to increase with LTE-U. 
	
	Additionally, when $\frac{\gamma B}{1-\alpha}<\frac{4}{3}$, we have 
	\begin{eqnarray}
	\label{eqn:revenue_gamma}
	\frac{\partial p_1x_1}{\partial \gamma} &=& \frac{\partial p_1x_1}{\partial W_e} \frac{\partial W_e}{\partial \gamma} + \frac{\partial p_1x_1}{\partial B_e}  \frac{\partial B_e}{\partial \gamma}\nonumber \\
	&=&\frac{\partial p_1x_1}{\partial B_e}  *\left[B+ \frac{\alpha\beta W}{1+\beta(1-\alpha)W/B}\right]\nonumber\\
	 &\ge& 0.\nonumber
	\end{eqnarray}
	Therefore, incumbent's revenue always increases with $\gamma$ when $\frac{\gamma B}{1-\alpha}<\frac{4}{3}$.
\end{proof}

\section{Proof of Theorem \ref{thm:one_one_sw_cs_spectral}}
\label{app:thm:one_one_sw_cs_spectral}
\begin{proof}
	We first show that the total customer mass $x_1+x_2$ increases with $\gamma$. We have 
	\begin{eqnarray}
	\label{eqn:x1x2_gamma}
	\frac{\partial x_1+x_2}{\partial \gamma} &=& \frac{\partial x_1+x_2}{\partial W_e} \frac{\partial W_e}{\partial \gamma} + \frac{\partial x_1+x_2}{\partial B_e}  \frac{\partial B_e}{\partial \gamma}\nonumber \\
	&=& \frac{\partial x_1+x_2}{\partial B_e}  *\left[B+ \frac{\alpha\beta W}{1+\beta(1-\alpha)W/B}\right].\nonumber
	\end{eqnarray}
	We also have 
		\begin{footnotesize}
		\begin{eqnarray}
		\label{eqn:d_x1+x2_Be}
		\frac{\partial (x_1+x_2)}{\partial B_e} &=&\frac{2 (1 + W_e)^2 (4 + 4 W_e+ 3 W_e^2)}{(1 + B_e + W_e)^2 (4 + 4 B_e + 
			4 W_e + 3 B_e W_e)^2}\nonumber\\
		&&+\frac{ B_e^2 (8 + 16 W_e + 9 W_e^2)}{(1 + B_e + W_e)^2 (4 + 4 B_e + 
			4 W_e + 3 B_e W_e)^2}\nonumber\\
		&&+\frac{
			4 B_e (4 + 10 W_e + 9 W_e^2 + 3 W_e^3)}{(1 + B_e + W_e)^2 (4 + 4 B_e + 
			4 W_e + 3 B_e W_e)^2}\nonumber\\
		&\ge& 0\nonumber	
		\end{eqnarray}
	\end{footnotesize}
	Consequently, we conclude that customer surplus increases with $\gamma$.
	
	With linear case inverse demand function and congestion function, social welfare can be expressed as
	\begin{equation}
	\label{eqn:sw_linear}
	SW = p_1 x_1 + p_2 x_2 + \frac{(x1+x2)^2}{2}.
	\end{equation}
	Differentiating $SW$ with $\gamma$ yields
	\begin{eqnarray}
	\label{eqn:sw_diff}
		\frac{\partial SW}{\partial \gamma} &=& \frac{\partial SW}{\partial W_e} \frac{\partial W_e}{\partial \gamma} + \frac{\partial SW}{\partial B_e}  \frac{\partial B_e}{\partial \gamma}\nonumber\\
		&=&\frac{\partial SW}{\partial B_e}  *\left[B+ \frac{\alpha\beta W}{1+\beta(1-\alpha)W/B}\right].\nonumber
	\end{eqnarray}
	Additionally, we have 
	\begin{scriptsize}
			\begin{eqnarray}
		\label{eqn:sw_be}
		\frac{\partial SW}{\partial B_e} &=& \frac{4 (1 + W_e)^3 (4 + 8 W_e + 3 W_e^2 + 2 W_e^3)}{((1 + B_e + W_e)^3 (4 + 4 B_e + 4 W_e + 3 B_e W_e)^3}\nonumber\\
		&&+\frac{2 B_e^4 (16 + 42 W_e + 35 W_e^2 + 9 W_e^3)}{((1 + B_e + W_e)^3 (4 + 4 B_e + 4 W_e + 3 B_e W_e)^3}\nonumber\\
		&&+\frac{B_e (1 + W_e)^2 (80 + 188 W_e + 128 W_e^2 + 37 W_e^3 + 3 W_e^4)}{((1 + B_e + W_e)^3 (4 + 4 B_e + 4 W_e + 3 B_e W_e)^3}\nonumber\\
		&&+\frac{B_e^3 (112 + 356 W_e + 398 W_e^2 + 179 W_e^3 + 24 W_e^4)}{((1 + B_e + W_e)^3 (4 + 4 B_e + 4 W_e + 3 B_e W_e)^3}\nonumber\\
		&&+\frac{3 B_e^2 (48 + 180 W_e + 252 W_e^2 + 161 W_e^3 + 44 W_e^4 + 3 W_e^5)}{((1 + B_e + W_e)^3 (4 + 4 B_e + 4 W_e + 3 B_e W_e)^3}.\nonumber
		\end{eqnarray}
	\end{scriptsize}
	 We can see that all terms in $\frac{\partial SW}{\partial B_e}$ are positive. Therefore,  social welfare increases with $\gamma$ as well.
\end{proof}

\section{Proof of Lemma \ref{lemma:duty_cycle_band}}
\label{app:duty_cycle_band}
	\begin{proof}
	The first statement can be proved by taking derivative of (\ref{eqn:Be}) and (\ref{eqn:We}) with respect to $\alpha$. For the second one, we need to calculate the derivative of $B_e+W_e$ with respect to $\alpha$. When $W>\frac{B}{1-\beta}$, we have the following inequality
	\begin{equation}
	\label{eqn:diff_b_w}
	\frac{\partial (B_e+W_e)}{\partial \alpha} < \frac{\beta^2(1+W/B)W[2\alpha-1]}{[1+(1-\alpha)\beta W/B]^2[1-(1-\alpha)\beta]^2}.\nonumber
	\end{equation}
	It implies that when $\alpha<\frac{1}{2}$, $\frac{\partial (B_e+W_e)}{\partial \alpha} <0$. As a result, $B_e+W_e$ will decrease with $\alpha$ in $(0,\frac{1}{2})$.
	\end{proof}

\section{Proof of Theorem \ref{thm:dutycycle}}
\label{app:thm:dutycycle}
	\begin{proof}
	According to Lemma \ref{lemma:duty_cycle_band} , $B_e$ increases with $\alpha$ and $W_e$ decreases with $\alpha$. When $\alpha$ increases, if the incumbent still announces the same price it is able to attract more consumers due to the decrease on congestion on licensed band. When there are multiple entrants, the entrants' strategy is announcing 0 price regardless of the bandwidth of licensed and unlicensed band. In this case, the incumbent can still adjust its price to gain more revenue.	That concludes our proof.
\end{proof}

\section{Proof of Theorem \ref{thm:optimal_alpha}}
\label{app:thm:optimal_alpha}
	\begin{proof}
	We first use best response function to calculate the equilibrium price and customer mass with $W\to\infty$. The results are as following
	\begin{eqnarray}
	\label{eqn:one_one_equilibrium_W_infty}
	p_1 = \frac{1}{4+3B_e},x_1 = \frac{B_e}{4+3B_e},\nonumber\\
	p_2 =\frac{2}{4+3B_e},x_2 = \frac{1+B_e}{4+3B_e}.\nonumber
	\end{eqnarray}
	We can find the revenue of incumbent 
	\begin{eqnarray}	
	p_1x_1& =&  \frac{B_e}{(4+3B_e)^2}\nonumber\\
	\label{eqn:one_one_profit}
	&=& \frac{1}{\frac{16}{B_e}+ 9B_e +24}
	\end{eqnarray}
	We can see that (\ref{eqn:one_one_profit}) increases in the interval $[0,\frac{4}{3}]$ and reaches maximum when $B_e = \frac{4}{3}$. We also have $\lim\limits_{W\to\infty} B_e = \frac{B}{1-\alpha}$. As a result, the optimal $\alpha = \max\{1-\frac{3B}{4},0\}$.
\end{proof}

\section{Proof of Theorem \ref{thm:sw_gap}}
\label{app:thm:sw_gap}
	\begin{proof}
	In this case, we have already shown that social welfare increases with $B_e$ in the proof of Theorem \ref{thm:one_one_sw}. The social welfare gap can be denoted as $Gap = SW(B_e) - SW(B)$. Since $\lim\limits_{W\to\infty} B_e = \frac{B}{1-\alpha}$, we can conclude that the social welfare gap is non-decreasing with $\alpha$.
\end{proof}

\section{Proof of Theorem \ref{thm:fix_uti_profit}}
\label{app:thm:fix_uti_profit}
	\begin{proof}
	Because $\alpha \beta = k$, we can replace $k/\alpha$ by $\beta$ in (\ref{eqn:equivalant_bw}). In order to prove the statement 1), we just need to prove revenue increases with $\beta$ under the given condition. We first find the derivative of the revenue $p_1x_1$ with respect to $\beta$. 
	\begin{equation}
	\label{eqn:profit_beta}
	\frac{\partial p_1x_1}{\partial \beta} = \frac{\frac{\partial B_e}{\partial \beta}(1+2W_e+W_e^2)-\frac{\partial W_e}{\partial \beta}(B_e^2+2B_e+2B_eW_e)}{4(1+W_e)^2(1+B_e+W_e)^2}. \nonumber
	\end{equation}
	For statement 1), suppose $W\le B$, we can verify that $\frac{\partial W_e}{\partial \beta}\le \frac{\partial B_e}{\partial \beta}<0$. Thus we can find the sufficient condition for the derivative of revenue to be non-negative.
	\begin{eqnarray}
	\label{eqn:profit_beta_sufficient}
	&&(B_e^2+2B_e+2B_eW_e)- (W_e+1)^2 \nonumber\\
	&=& (1+B_e+W_e)^2-2(1+W_e)^2 \ge 0 \nonumber
	\end{eqnarray}
	By solving the inequality, we can find the sufficient condition for the revenue to increase with $\beta \in (k,1)$ is $B\ge\frac{\sqrt{2}}{2}$ and $W\le B$. 
	
	Then we prove statement 2). We only need to find the condition such that $\lim \limits_{\beta\to k} \frac{\partial p_1x_1}{\partial \beta} < 0$.  We have
	\begin{eqnarray}
	\label{eqn:lim_beta_to_k}
	\lim\limits_{\beta\to k} \frac{\partial p_1x_1}{\partial \beta}& = &  C_0\left[(B+kW)(B+kW+2(1-k)W+2) \right.\nonumber\\
	&&-\left.[1+(1-k)W]^2W/B \right], \nonumber
	\end{eqnarray}
	where $C_0$ is a positive component. We can as a result find the sufficient condition
	\begin{equation}
	\label{eqn:sufficient2}
	\frac{W}{B}> \frac{(B+kW)(B+kW+2(1-k)W+2) }{[1+(1-k)W]^2}
	\end{equation}
	Note that for any choice of $B$, LHS of (\ref{eqn:sufficient2}) is increasing linearly with respect to $W$ and RHS will increase to some constant when $W$ is large, because the most significant parts of the numerator and denominator are $W^2$. As a result, there must exist some $W_{th}$ such that when $W>W_{th}$, inequality (\ref{eqn:sufficient2}) holds. Then we can reach the conclusion.
\end{proof}
\end{document}